\pdfoutput=1
\documentclass[11pt,a4paper]{article}
\usepackage{jcappub}
\usepackage{amsmath}
\usepackage{mathtools}
\usepackage{graphicx}
\usepackage{longtable}
\usepackage{bm}
\usepackage{amsfonts}
\usepackage{subfigure}
\usepackage{breqn}
\usepackage{multirow}
\usepackage{float}
\usepackage{wrapfig}
\usepackage{array}
\usepackage{multimedia}

\long\def\dump#1{}

\usepackage{xcolor}

\begin{document}


\title{
High energy particles from young supernovae:  gamma-ray and  neutrino connections}

\author[a]{Prantik~Sarmah,}
\author[a]{Sovan~Chakraborty,}
\author[b]{Irene~Tamborra,}
\author[c,d,e]{and Katie~Auchettl}

\affiliation[a]{Department of Physics, Indian Institute of Technology Guwahati,
Guwahati, Assam-781039, India}

\affiliation[b]{Niels Bohr International Academy and DARK, Niels Bohr Institute, University of Copenhagen, Blegdamsvej 17, 2100, Copenhagen, Denmark}

\affiliation[c]{School of Physics, The University of Melbourne, Parkville, VIC 3010, Australia}
\affiliation[d]{ARC Centre of Excellence for All Sky Astrophysics in 3 Dimensions (ASTRO 3D)}
\affiliation[e]{Department of Astronomy and Astrophysics, University of California, Santa Cruz, CA 95064, USA}

\emailAdd{prantik@iitg.ac.in}
\emailAdd{sovan@iitg.ac.in}
\emailAdd{tamborra@nbi.ku.dk}
\emailAdd{katie.auchettl@unimelb.edu.au}

\abstract{
Young core-collapse supernovae (YSNe)
are  factories of high-energy neutrinos and gamma-rays as the shock accelerated protons efficiently interact with the protons in the 
dense circumstellar medium. We explore the detection prospects of secondary particles from YSNe of Type 
IIn, II-P, IIb/II-L, and Ib/c. 
Type IIn YSNe are found to produce the largest  flux of neutrinos and gamma-rays, followed by II-P YSNe. 
 Fermi-LAT and the Cherenkov Telescope Array (IceCube-Gen2) have the potential to detect Type IIn YSNe  up to $10$~Mpc ($4$~Mpc), 
 with the remaining YSNe Types being detectable  closer to Earth.  
 We also find that  YSNe may dominate the diffuse neutrino background, especially between $10$~TeV and $10^3$~TeV, while they do not constitute a dominant component to the isotropic gamma-ray background observed by Fermi-LAT. At the same time, the IceCube high-energy starting events  and Fermi-LAT data already allow us to exclude a large fraction of the model parameter space of YSNe otherwise inferred from multi-wavelength electromagnetic observations of these transients. 
}

\maketitle

\section{Introduction}
\label{sec:introduction}
The IceCube Neutrino Observatory routinely detects high-energy neutrinos of astrophysical origin, whose origin remains unknown \cite{IceCube:2018fhm,IceCube:2021xar,IceCube:2020wum,IceCube:2020acn}. Various source classes have been proposed to explain the IceCube neutrino flux~\cite{Meszaros:2017fcs,Ahlers:2018fkn,Vitagliano:2019yzm,Kurahashi:2022utm}, such as galaxy clusters, (low-luminosity or chocked) gamma-ray bursts, tidal distruption events,  star-forming galaxies, and supernovae (SNe)~\cite{Meszaros:2015krr,Pitik:2021xhb,Murase:2015ndr,Waxman:2015ues,Tamborra:2014xia,Zandanel:2014pva,Wang:2015mmh,Dai:2016gtz,Senno:2016bso,Lunardini:2016xwi,Petropoulou:2017ymv,Tamborra:2015fzv,Denton:2017jwk}. However, none of these sources can fully explain the IceCube data, hinting that more than one  source class contributes to the overall observed flux~\cite{Palladino:2018evm,Bartos:2021tok}.  In addition to the diffuse background of high-energy neutrinos,   neutrino events have been observed in likely associations with blazars~\cite{IceCube:2018dnn,Giommi:2020viy,Franckowiak:2020qrq,Fermi-LAT:2019hte,Krauss:2018tpa,Kadler:2016ygj}, tidal distruption events~\cite{Stein:2020xhk,Reusch:2021ztx}, and a hydrogen-rich super-luminous SN~\cite{Pitik:2021dyf}. Moreover, ongoing electromagnetic follow-up searches,  see e.g.~\cite{IceCube:2020mzw,Fermi-LAT:2019hte,VERITAS:2021mjg}, promise to provide key insight to  disentangle the origin of the ever growing number of IceCube neutrino events of astrophysical origin.

The high-energy neutrino events observed by the IceCube Neutrino Observatory  are believed to have a correspondent counterpart in gamma-rays possibly observable by Fermi-LAT, see e.g.~\cite{Capanema:2020oet,Tamborra:2014xia,Ando:2015bva,Feyereisen:2016fzb,Krauss:2018tpa,Peretti:2019vsj,Chakraborty:2016mvc,Hooper:2016jls,Sudoh:2018ana,Murase:2013rfa,Wang:2019bcs}. This is because such high energy particles  originate from the decay of charged and neutral pions produced in hadronic interactions such as proton-proton ($pp$) collisions or photo-hadronic ($p\gamma$) interactions~\cite{Kelner:2006tc,Kelner:2008ke}. In this respect special attention has been devoted to the possible common origin of the IceCube diffuse flux of high-energy neutrinos~\cite{Abbasi:2020jmh} and the extragalactic diffuse gamma-ray background observed by Fermi-LAT~\cite{Fermi-LAT:2015otn}. Evidence suggests that the latter  originates from the superposition of unresolved extragalactic sources, such as blazars, star-forming galaxies, gamma-ray bursts,  etc.~\cite{Dermer:2007fg,Inoue:2011bp,Fermi-LAT:2010tsy,Fermi-LAT:2012nqz,Stecker:2010di,DiMauro:2013zfa,Ajello:2020zna,Lisanti:2016jub,Fermi-LAT:2015otn,Massaro:2015gla,Fornasa:2015qua,Tamborra:2014xia,Bechtol:2015uqb,Roth:2021lvk}.

Supernovae constitute a particularly interesting source class, possibly emitting high-energy neutrinos and gamma-rays through  inelastic $pp$ collisions between the relativistic protons accelerated at the SN shock and the low energy  protons of the circumstellar medium (CSM)~\cite{Halzen:2002pg, Becker:2007sv,Murase:2019tjj, Tamborra:2018upn,Chevalier:2001at,Murase:2010cu,Murase:2013kda}. In fact, the existence of dense CSM around massive stars has been confirmed by multi-wavelength observations of a wide range of phenomena, e.g.,~\cite{Smith:2014txa,2011BSRSL..80..322S}. The CSM   is made up of matter deposited through stellar winds further    enriched  as the progenitor star loses mass via wind and/or violent outbursts (e.g., see Ref.~\cite{Smith:2014txa} and references therein). The CSM was previously thought to originate from  stars losing their mass through steady stellar winds.
However, recent SN observations  have challenged this traditionally accepted picture  (see e.g. \cite{Smith:2014txa}). It is now clear that substantial and impulsive mass losses occur in at least $10\%$ of massive stars within one year from the  core collapse. 

The  stellar progenitor of SN2009ip, whose  repeated eruptions occurred in the years before explosion, is the best studied example among Hydrogen (H)-rich massive 
stars~\cite{Margutti:2013pfa}. 
The progenitor of this SN was a Luminous Blue Variable (LBV) star. An LBV is a massive star that can have sporadic, violent mass loss events and  exhibits mass loss rate as high as $10$ $ \rm M_{\odot} yr^{-1}$ \cite{Humphreys:1989yqb,doi:10.1098/rsta.2016.0268,2020Galax...8...20W,Groh:2013gy}. 
LBVs were previously thought to be a transitional phase of stellar evolution. However, several recent works have shown that they can be progenitors of   Type IIn SNe (see e.g.~\cite{Ustamujic:2021zuv} and references therein). The interaction between the CSM and the SN ejecta is evident from the observation of narrow, symmetric emission lines in ``flash" spectroscopy of  SNe. This might be due to the  photo-ionization of the CSM occurring during the progenitor mass loss phase before explosion, see e.g.~\cite{Jacobson-Galan:2021pki,Terreran:2021hfc}. UV/optical excesses observed in SNe at early times also hint towards such mass loss phenomena, see e.g.~\cite{YoungSupernovaExperiment:2021fur}. 

In most cases, the interaction of the SN shock with the CSM is  dominant during the early stages of the SN evolution,  for a few months to a few years~\cite{Ofek:2013afa}. After this preliminary phase, the interaction weakens due to the  fall in the CSM density~\cite{Chandra:2017aev}. We focus on the high-energy particle emission during this initial SN phase and refer to it as ``Young Supernova (YSN)''.
Note that, a clear distinction between YSNe and  SN remnants (SNRs) is not straightforward \cite{Sturner_1997,Chandra:2008aq,Chevalier:2001at}. In a SNR, the SN ejecta interacts with and sweeps up the far away CSM or  the Interstellar Medium (ISM). For a sufficiently old SNR ($1~\rm kyr$), the medium is significantly less dense and has a density of $1-10^3$ $\rm cm^{-3}$. While the dense CSM during the YSN phase can have  densities in the scale of $10^{9}-10^{12}$ $\rm cm^{-3}$~\cite{Roth:2021lvk,Chakraborty:2015sta,Petropoulou:2016zar,Petropoulou:2017ymv}. Thus the largest contribution to the secondary particle emission is reasonably considered to originate  from the YSN phase.

The high-energy neutrinos  emitted from SNe are  possibly detectable by the IceCube Neutrino Observatory~\cite{Murase:2017pfe,Petropoulou:2017ymv,Chakraborty:2016mvc,Tamborra:2018upn}. 
In addition, gamma-rays could be observed from SNe  by Fermi-LAT~\cite{Fermi-LAT:Sensitivity,Fermi-LAT:2010tsy,Fermi-LAT:2015otn} and the Cerenkov Telescope Array (CTA)~\cite{2011ExA....32..193A}. 
Recently, Fermi-LAT  detected gamma-rays from the direction of a peculiar supernova iPTF14hls \cite{Yuan:2017oqu}. Such discovery is however  uncertain because of the presence of a blazar in the detection error circle. 
 
Varying mass and metallicity of stars may give rise to different SN Types~\cite{Gal-Yam:2016yms,Turatto:2003np}. By considering YSNe of Type IIn, II-P, II-L and Ib/c, we explore the related gamma-ray and neutrino emission  and investigate their detection prospects with Fermi-LAT~\cite{Fermi-LAT:Sensitivity,Fermi-LAT:2015otn} and CTA~\cite{2011ExA....32..193A} as well as  IceCube~\cite{IceCube:2016tpw,IceCube:2020wum}, IceCube-Gen2~\cite{IceCube-Gen2:2020qha}, and Km3NeT~\cite{KM3NeT:2018wnd}.

While neutrinos produced in YSNe propagate undisturbed to Earth,  gamma-rays  undergo energy losses as they interact with  low energy photons ($\gamma\gamma\longrightarrow e^{-}e^{+}$) and ambient matter ($\gamma N\longrightarrow N e^{-}e^{+}$, where $N$ is nucleus) in the source~\cite{Murase:2018okz}. 
In addition, over cosmic scales ($\sim$Gpc), the Extra-galactic Background Light (EBL) absorption further affects the gamma-ray flux expected at Earth~\cite{Stecker:2016fsg,Stecker:2010di}, becoming  significant for the diffuse SN emission, but negligible for point source detection in the local universe  (i.e., below $\sim 10$ Mpc). 

This paper is organized as follows.  In Sec.~\ref{sec:SNmodel}, we introduce our YSN model with interacting CSM and  the related production of secondary particles: neutrinos and gamma-rays. Different Types of YSNe and their properties are discussed in Sec.~\ref{sec:DOSNR}. The dependence of the gamma-ray and neutrino emission on the  YSN Types is explored in  Sec.~\ref{sec:point sources}. The diffuse  gamma-ray and neutrino backgrounds from YSNe are presented in Sec.~\ref{sec:diffuse} together with a discussion on the model parameter uncertainties and the detection prospects. The  detection prospects of YSNe in the local universe with present and upcoming gamma-ray and neutrino telescopes are analysed in Sec.~\ref{sec:Point_source_detection}. 
Finally, we summarize our findings  in Sec.~\ref{sec:conclusion}. An overview on the characteristic time scales for particle acceleration and energy losses of protons is provided in Appendix~\ref{sec:appendix}.

\section{Neutrino and gamma-ray production in young supernovae}\label{sec:SNmodel}
In this Section, we model YSNe  interacting with the CSM in the early phase after the explosion. We focus on YSNe  about a year old
and develop a model for the shock-CSM interaction, describing the creation of  secondary particles: neutrinos and anti-neutrinos~\footnote{Note that we do not distinguish between particles and antiparticles in the following and use  ``neutrinos'' to indicate both species.} and gamma-rays. 

\subsection{Model setup}
\label{sec:Model}
    The material expelled by the massive star at the end of its life forms the CSM.
    The CSM consists of shells of mostly  light elements like H or helium (He). The massive star eventually explodes in the form of a core-collapse SN~\footnote{We focus on core-collapse SNe as the vast majority of Type Ia SNe do not exhibit a dense CSM (with the notable exception of e.g., \cite{Fox:2014zlm}).}. 
    The SN  shock and the expanding ejecta interact with the surrounding CSM. Charged particles (electrons and protons) in the shocked CSM are accelerated to relativistic energies via Fermi's diffusive shock acceleration~\cite{1997ApJ...490..619S,Bell:2013gga,gaisser_engel_resconi_2016}. In particular, we are interested in high energy processes with dominant neutrino production, hence we focus on  inelastic $pp$ collisions. Here, relativistic protons collide with the non-relativistic protons (inelastic $pp$ collisions) in the CSM and  produce a large number of $\pi$ and $\eta$ mesons. Secondary neutrinos, 
    electrons and gamma-rays are produced from the decay of  $\pi$ and $\eta$ mesons~\cite{1982ApJ...258..790C,Mastichiadis:1995pz,Kelner:2006tc, Petropoulou:2016zar,Ofek:2014fua,Petropoulou:2017ymv}. High energy protons can also undergo photo-hadronic ($p\gamma$) interactions with low energy photons in the CSM. However, the energy loss due to this process is negligble and hence neglected (see Appendix \ref{appendix}).

The CSM shell is assumed to be spherically symmetric with an effective inner radius $r_{\rm in}$ and outer radius $r_{\rm out}$. The inner radius is defined as $r_{\rm in}= \max[r_{\rm bo},r_{\rm e}]$ \cite{Murase:2017pfe}. The shock breakout radius\footnote{The shock breakout radius corresponds to the optical depth, $\tau =c/v_{\rm sh}$, of the material through which the shock propagates, where $v_{\rm sh}$ is the shock velocity introduced in the following \cite{Waxman2017}.}, $r_{\rm bo}$, corresponds to the beginning of  shock acceleration, and $r_{\rm e}$ is the radius of the stellar envelope. Thus $r_{\rm in}$ may differ according to the SN Type (see Sec.~\ref{sec:DOSNR} for details).  
The outer radius or size of the CSM shell also depends on the SN Type. Given the assumption of spherical symmetry, the radial dependence of the CSM mass density can be modelled in the following way~\cite{Chevalier:2001at}:
\begin{equation}
    \rho_{\rm CSM}(r) = \frac{\dot{M}_{\rm W}}{4 \pi r_{\rm in}^w v_{\rm w}} \left(\frac{r_{\rm in}}{r}\right)^w.
    \label{eq:CSM_density}
\end{equation}
Here $w$ is the power-law index of the CSM density profile (e.g., $w=2$ for a 
wind-like CSM---see, e.g., Ref.~\cite{Ofek:2013afa}), $\dot{M}_{\rm W}$ is the mass loss rate of the progenitor star and $v_{\rm w}$ is the wind velocity.
The radial dependence is provided by the propagation of the shock, $r$ being the shock radius. The number density of the protons  in the CSM ($n_{\rm CSM}(r)= \rho_{\rm CSM}(r)/m_{\rm p}$) also varies accordingly. Thus, the CSM at the inner radius is $n_{\rm in,CSM}= {\rho_{\rm CSM} (r_{\rm in})}/{m_p}={\dot{M}_{\rm W}}/{(4 \pi m_p r_{\rm in}^2 v_{\rm w})} $, where $m_p$ is the mass of the proton.

A fraction of the protons injected in the shocked region will be accelerated to high energies and in turn produce secondary particles while interacting with the CSM protons. To model this, the SN shock is assumed to expand spherically in the CSM  and has a power law dependence on the radius. The velocity of the shock ($v_{\rm sh}$) expanding in the CSM varies slowly with radius~\cite{Ofek:2013afa}:
\begin{equation}
  v_{\rm sh} (r) = v_{\rm in} \left(\frac{r}{r_{\rm in}}\right)^\alpha\ ,  
  \label{proton_model}
\end{equation}
where $v_{\rm in}$ is the shock velocity at $r_{\rm in}$ and $\alpha$ characterises the shock velocity profile. The index $\alpha$ can be expressed in terms of the power law indices of the CSM and SN ejecta density profiles and is usually found to be negligible, i.e.~the shock velocity is nearly constant over the length of the CSM shell during the early phase of the YSN. For more details on the choice of $\alpha$, see Ref.~\cite{Ofek:2014fua}. In our YSN computations, we assume $\alpha=0$ implying $v_{\rm sh} = v_{\rm in}$.

To estimate the secondary particle population, we need the injection spectra of the accelerated protons in the shocked region.  We assume a power law distribution between the maximum ($E_{\rm p, max}$) and minimum ($E_{\rm p, min}$) proton energy for the injection spectra~\cite{Petropoulou:2016zar}: 
\begin{equation}
    N_{\rm p}^{\rm inj}(E_{\rm p},r) \propto E_{\rm p}^{-\alpha_{\rm p}}  \exp\left(-\frac{E_{\rm p}}{E_{\rm p,max}(r)} \right)\ , 
    \label{eq:proton_source}
\end{equation}where $E_{\rm p}$ is the proton energy, ${\alpha_{\rm p}}$ is the power law index and it is assumed to be equal to  2 unless otherwise mentioned. The choice ${\alpha_{\rm p}}=2$ is motivated by diffusive shock acceleration theory~\cite{gaisser_engel_resconi_2016}. 
The normalization of the injected spectra is obtained from the energy budget of the shock [$E_{\rm K}= (9 \pi/8) m_{\rm p} v_{\rm sh}^2 r^2 n_{\rm CSM}(r)$, energy per unit radius]. A fraction of this energy, $\epsilon_{\rm p}$, is transferred to the charged particles (protons) confined in the shocked region due to the presence of a strong magnetic field and thus accelerates protons to high energies.
The fraction $\epsilon_{\rm p}$ 
is a free parameter of the problem and may depend on the SN Type. This is due to the fact that CSM density, ejecta kinetic energy and progenitor mass vary for different SN Types~\cite{Caprioli_2014}.  The minimum proton energy $E_{\rm p,min}$ is taken as the proton rest mass $m_{\rm p}$.  The maximum energy ($E_{\rm p,max}$) of the accelerated protons depends on the different loss processes competing with the acceleration time scale ($t_{\rm acc}= 6 E_{\rm p} c/ e B v_{\rm sh}^2$) \cite{Protheroe:2003vc,Petropoulou:2016zar}, $B$ being the magnetic field strength. The magnetic field is   inversely proportional to the shock radius $r$ and it is $B=3/2 (4 \pi \epsilon_{\rm B} m_{\rm p} n_{\rm in,CSM} v_{\rm sh}^2)^{1/2} (r_{\rm in}/r)$, where $\epsilon_{\rm B}$ is the fraction of the post shock thermal energy that goes into the magnetic field~\cite{Petropoulou:2016zar,RevModPhys.76.1143,Meszaros:2006rc}. A detailed analysis on the effects of magnetic field on $E_{\rm p,max}$ can be found in Ref.~\cite{Marcowith:2018ifh}.

The energy losses suffered by relativistic protons mainly come from  two competing processes, due to the adiabatic expansion of the shock shell and  the $pp$ interactions with the CSM protons.  The adiabatic loss time scale or the dynamical time scale is $t_{\rm ad}\sim t_{\rm dyn} = r/v_{\rm sh} $ and the $pp$ collision time scale is $ t_{\rm pp}=[4\kappa_{\rm pp} \sigma_{\rm pp} n_{\rm CSM}(r) c]^{-1}$, where $\kappa_{\rm pp}=0.5$ is the proton inelasticity and $\sigma_{pp}$ is the $pp$ interaction cross-section. We assume $\sigma_{\rm pp}\approx3\times 10^{-26}$ $\rm cm^2$, neglecting the  energy dependence of $\sigma_{\rm pp}$ since it hardly affects 
$t_{\rm pp}$~\cite{Petropoulou:2016zar,Petropoulou:2017ymv} (however, as specified later, we include the energy dependence in $\sigma_{\rm pp}$ for the calculation of the secondaries). In addition, protons undergo other energy loss processes, as detailed
in Appendix \ref{sec:appendix}.   
Thus, $E_{\rm p,max}$ can be estimated from  $t_{\rm acc}=6 E_{\rm p,max} c/ e B v_{\rm sh}^2=\mathrm{min}[t_{\rm pp},t_{\rm ad}]$ and changes according to the radial evolution of the background environment.

The radial evolution of the steady state proton number, $N_{\rm p}(E_{\rm p},r)$ is described by the following equation~\cite{Mastichiadis:1994hq,1997ApJ...490..619S,2012ApJ...751...65F,Petropoulou:2016zar,Petropoulou:2017ymv}:
\begin{equation}
\frac{\partial N_{\rm p}(E_{\rm p},r)}{\partial r}+\frac{N_{\rm p}(E_{\rm p},r)}{v_{\rm sh} t_{\rm pp}(r)}-\frac{\partial}{\partial E_{\rm p}}\left[\frac{E_{\rm p} N_{\rm p}(E_{\rm p},r)}{r}\right]= N_{\rm p}^{\rm inj}(E_{\rm p},r)\ ,
\label{proton_de}
\end{equation}
where the second term on the left hand side corresponds to energy losses via $pp$ collisions  and the third term stands for adiabatic losses. This primary steady state proton number drives the production of the secondary neutrinos and gamma-rays. The time evolution of the fluxes can be probed from the relation betweeen the shock radius and the shock velocity:  $r=v_{\rm sh} t$.

\subsection{Gamma-ray production} 
\label{subsec:source abs}
Gamma-rays can  be produced through $pp$ interactions via decay of neutral $\pi$ and $\eta$ mesons, with the following decay channels: $\pi^{0} \longrightarrow 2 \gamma$, $\eta \longrightarrow 2\gamma$, $\eta \longrightarrow 3 \pi^0$, $\eta \longrightarrow \pi^+\pi^-\pi^0$, and $\eta \longrightarrow \pi^+ \pi^{-} \gamma$ \cite{Kelner:2006tc}.  The radial evolution of gamma-rays with energy $E_{\gamma}$ produced in $pp$ collisions can be computed as follows~\cite{Murase:2018okz,Murase:2014bfa}:
\begin{equation}
    \frac{\mathrm{d}N_{\gamma}(E_{\gamma},r)}{\mathrm{d}r} + \frac{N_{\gamma}(E_{\gamma},r)}{v_{\rm sh} t_{\rm esc}}  = N^{\rm inj}_{\gamma} (E_{\gamma},r),
    \label{gamma_de}
\end{equation}
where $t_{\rm esc}=r/4 c$ is the escape time of secondary particles. The factor $4$  corresponds to the compression of the CSM due to shock pressure \cite{1997ApJ...490..619S}.   The source term, $N^{\rm inj}_{\gamma}(E_{\gamma},r)$, below $E_{\rm p}=0.1$ TeV is given by \cite{Kelner:2006tc} 
\begin{equation}
     N^{\rm inj}_{\gamma}(E_{\gamma},r) = 2\frac{ \tilde{\eta}}{K_{\pi}}  \frac{4 c n_{\rm CSM}(r) m_{\rm e}}{v_{\rm sh} m_{\rm p}} \int_{E_{\rm min}}^{\infty} \frac{  \mathrm{d}E_{\pi}}{\sqrt{E_{\pi}^2-m_{\pi}^2}} \sigma_{\rm pp} \left( E_{\rm p} \right) N_{\rm p}\left( E_{\rm p} \right)\ ,
\end{equation} where $E_{\rm p}=m_{\rm p} + {E_{\pi}}/{K_{\pi}} $ and $E_{\rm min} = E_{\gamma} + {m_{\pi}^2}/{E_{\gamma}}$. Above $E_{\rm p}=0.1$ TeV,
\begin{equation}
    N^{\rm inj}_{\gamma}(E_{\gamma},r) =\frac{4 c n_{\rm CSM}(r) m_{\rm e}}{v_{\rm sh} m_{\rm p}}\int_{m_{\rm p}}^{\infty} \mathrm{d}E_{\rm p} \frac{\sigma_{\rm pp}(E_{\rm p})}{E_{\rm p}} N_{\rm p}(E_{\rm p},r) F_{\gamma}\left(\frac{E_{\gamma}}{E_{\rm p}}\right)\ ,
\end{equation}
with $F_{\gamma}\left({E_{\gamma}}/{E_{\rm p}}\right)$ being the gamma-ray production rate,   and $E_{\gamma}$ the  gamma-ray energy. The parameters $K_{\pi}$ and $\tilde{\eta}$ are free parameters, used for connecting the injection rates above energy $E_{\rm p}= 1$ TeV.

The total gamma-ray flux observed at Earth from a source at luminosity  distance $d_{\rm L}$  is given by: 
\begin{align}
     E_{\gamma,\rm obs}^2 \Phi_{\gamma}(E_{\gamma,\rm obs}) &=\frac{  e^{-\tau_{\gamma,\rm EBL}(E_{\gamma},z)} }{4 \pi d_{\rm L}^2}  E_{\gamma}^2 \phi_{\gamma}^{\rm s}(E_{\gamma})\nonumber \\&= \frac{e^{-\tau_{\gamma,\rm EBL}(E_{\gamma},z)}}{4 \pi d_{\rm L}^2} \int_{r_{\rm in}}^{r_{\rm max}} \mathrm{d}r \frac{ E_{\gamma}^2  N_{\gamma}(E_{\gamma},r) e^{-\left( \tau_{\gamma\gamma}+\tau_{\rm BH}\right)}}{ m_{\rm e} c^2 t_{\rm esc} v_{\rm sh}}\ ,
     \label{eq:gamma_at_earth}
\end{align}
where $E_{\gamma} = (1+z)E_{\gamma,\rm obs}$, $z$ is the cosmological redshift and $\phi^{\rm s}_{\gamma}$ is the gamma-ray flux at source. The maximum shock radius, $r_{\rm max}$ corresponds to the end of particle production (see Sec. \ref{subsec:neutrinos} for details). Here, $\tau_{\gamma\gamma}$ and $\tau_{\rm BH}$ are the optical depths for gamma-gamma  and Bethe-Heitler interactions, respectively; their modeling is introduced in the following.  $\tau_{\gamma, \rm EBL}(E_{\gamma},z)$ is the optical depth for the interaction of gamma-rays with the EBL~\cite{Stecker:2005qs,Stecker:2016fsg}.

The gamma-rays produced in the hadronic processes can interact with the low energy thermal photons present in the CSM \cite{Murase:2018okz,Murase:2014bfa,Petropoulou:2016zar}. In these  interactions,  electron-positron pairs  can be produced.  Electron-positron pairs can also be produced in Bethe-Heitler processes, where  high-energy photons interact with nuclei (mostly protons)  in the ejecta~\cite{Murase:2014bfa}.

The attenuation of the gamma-ray spectra due to photon-photon pair production is taken into account through the factor $e^{-\tau_{\gamma\gamma}}$ in Eq.~\ref{eq:gamma_at_earth};  the  optical depth $\tau_{\gamma\gamma}$ of the CSM thermal photons is computed as follows
\begin{equation}
\tau_{\gamma\gamma} = r_{\rm max}\int_0^{\infty} n_{\rm ph} (\epsilon) \sigma_{\gamma\gamma} (E_{\gamma},\epsilon) \mathrm{d} \epsilon\ ,     
\end{equation}
 where $\epsilon$ is the energy of thermal photons and  $\sigma_{\gamma\gamma}(E_{\gamma},\epsilon)$ is the photon-photon annihilation cross section~\cite{Jauch:1976ava,Murase:2014bfa}. The number density of thermal photons follows a black-body spectrum,
\begin{equation}
    n_{\rm ph} (\epsilon) \propto \frac{\epsilon^2}{\exp(\epsilon/ T)+1}\ ,
\end{equation}
where  $T$ is the temperature of the thermal photons and can be defined in terms of average energy, $\epsilon_{\rm av}$ of photons as $T=\epsilon_{\rm av}/3.15$. The constant of proportionality is such that 
\begin{equation}
    \int_0^{\infty} \mathrm{d}\epsilon  n_{\rm ph}(\epsilon) = U_{\rm ph, av}\ .
\end{equation}
The average energy density of  thermal photons in the CSM is:
\begin{equation}
    U_{\rm ph, av}=\frac{1}{r_{\rm max}}\int_{r_{\rm in}}^{r_{\rm max}} \mathrm{d}r \frac{ L_{\rm SN, pk}}{4 \pi c r^2}\ ,
\end{equation}
where  $L_{\rm SN, pk}$ is the SN peak luminosity.
In general, the energy density of  thermal photons scales as  $1/r^2$ with the shock radius, $r$.  We take the average energy density of the thermal photons in the interaction zone, i.e.~between $r_{\rm in}\leq r < r_{\rm max}$, to make a reasonable estimate of the absorption of  gamma-rays caused by the thermal photons.

In addition to $\gamma\gamma$ pair production losses, Bethe-Heitler pair production losses can also affect the gamma-ray spectra. The loss due to this process can be estimated by the factor $e^{-\tau_{BH}}$ in  Eq.~\ref{eq:gamma_at_earth}, where $\tau_{BH}$ is the optical depth. The latter  depends on the mass and composition of the ejecta and we model it as in Ref.~\cite{Murase:2014bfa}.  

 The electrons (positrons) produced in these processes may lose energy via synchrotron radiation due to the presence of magnetic fields in the shocked CSM. They also lose energy due to inverse Compton scattering with low-energy photons in the CSM. Electrons and positrons  can then annihilate and produce low-energy gamma-rays, which in turn modify the observed gamma-ray spectra.  This cascading of electromagnetic (EM) interactions modifying the gamma-ray spectrum can be estimated with a broken power law and is given by~\cite{Chang:2014hua},
\begin{equation}
    \phi_{\gamma,\rm cascade}(E_{\gamma})  \propto
    \begin{cases}
    E_{\gamma}^{-1.5} & (E_{\gamma}< E_{\gamma,\rm b}) \\
    E_{\gamma}^{-\alpha_{\gamma}} & (E_{\gamma,\rm b}< E_{\gamma}<E_{\rm cut}) 
    \end{cases}
    \label{eq:EMcascade}
\end{equation}
where  $E_{\rm cut}$ is the cut-off energy defined as $\tau_{\gamma\gamma}(E_{\gamma}=E_{\rm cut}=E_{\rm max})=1$, $E_{\gamma,\rm b}$ is the break energy given by $E_{\gamma,\rm b} = {4 E_{\rm cut}^2}/{3 (2 m_{\rm e} c^2)^2} \epsilon_{\rm av} $ and the power law index $\alpha_{\gamma}=2$.  This cascaded flux is  normalized to the total energy lost above $E_{\rm cut}$ due to pair production. This suggests that the larger the absorption, the   larger the cascaded flux.  Therefore, harder spectra would produce large cascaded gamma-ray flux.

The primary electrons accelerated in the shock could also produce gamma-rays via inverse Compton. The gamma-ray contribution of $pp$  collisions (i.e., $\pi^0 \longrightarrow 2 \gamma$) generally dominates in the high-energy range (GeV), hence  the gamma ray contribution from the primary electrons can be  ignored, see e.g.~\cite{Murase:2018okz}.


\subsection{Neutrino production}
\label{subsec:neutrinos}
Neutrinos  are produced through the decay of charged pions created in $pp$ interactions,  $\pi \longrightarrow \mu \nu_{ \mu}$ and $\mu \longrightarrow e \nu_{e} \nu_{\mu}$. The decay of $\eta$ mesons also produces neutrinos, but their contribution is smaller than the one of pions~\cite{Kelner:2006tc}.  The evolution  of neutrinos at different radii is governed by the following equation \cite{Petropoulou:2017ymv}:
\begin{equation}
    \frac{\mathrm{d} N_{\nu_{\rm f}} (E_{\nu},r)}{\mathrm{d}r} + \frac{N_{\nu_{\rm f}} (E_{\nu},r)}{v_{\rm sh} t_{\rm esc}(r)} = N^{\rm inj}_{\nu_{\rm f}}(E_{\nu},r)\ , 
\end{equation}
where $t_{\rm esc}(r)=r/ 4 c$ is the time required for neutrinos to escape the CSM (see comments below Eq.~\ref{gamma_de}).
of flavour  $ f$ (${ f}=e,\mu$ at the source) with neutrino energy $E_{\nu}$.  The neutrino injection term $N^{\rm inj}_{\nu_{\rm f}}(E_{\nu},r) $ is obtained by following Ref.~\cite{Kelner:2006tc}.

For $E_{\rm p}< 0.1$~TeV:
\begin{align}
    N^{\rm inj}_{\nu_{e}}(E_{\nu},r) &= 2\frac{ \tilde{\eta}}{K_{\pi}}  \frac{4 c n_{\rm CSM}(r) m_{\rm e}}{v_{\rm sh} m_{\rm p}} \int_{E_{\rm min}}^{\infty} \frac{  \mathrm{d}E_{\pi}}{\sqrt{E_{\pi}^2-m_{\pi}^2}} \sigma_{\rm pp} \left( E_{\rm p} \right) N_{\rm p}\left(E_{\rm p}, r \right) f_{\nu_e}\left(\frac{E_{\nu}}{E_{\pi}} \right) \nonumber \\
    N^{\rm inj}_{\nu_{\mu}}(E_{\nu},r) &= 2 \frac{ \tilde{\eta}}{K_{\pi}}  \frac{4 c n_{\rm CSM}(r) m_{\rm e}}{v_{\rm sh} m_{\rm p}} \int_{E_{\rm min}}^{\infty} \frac{  \mathrm{d}E_{\pi}}{\sqrt{E_{\pi}^2-m_{\pi}^2}} \sigma_{\rm pp} \left( E_{\rm p} \right) N_{\rm p}\left( E_{\rm p},r \right)& \nonumber \\ & \hspace{6 cm}\times
    \left[f_{\nu_{\mu}}^{(1)}\left(\frac{E_{\nu}}{E_{\pi}} \right)+f_{\nu_{\mu}}^{(2)}\left(\frac{E_{\nu}}{E_{\pi}} \right)\right]
      \label{nu_source1}
\end{align}
where $E_{\rm p}=m_{\rm p} + {E_{\pi}}/{K_{\pi}}$, and $E_{\pi}$ is the energy of pion and  $m_{\rm e}$ is mass of electron. The minimum energy of pion is given by  $E_{\rm min} \simeq E_{\nu}+ {m_{\pi}^2}/{4 E_{\nu}}$. The parameters $K_{\pi}$ and $\tilde{\eta}$ are free parameters and used for connecting the injection rates above energy $E_{\rm p}= 1$ TeV. 

For $E_{\rm p}\geq 0.1$~TeV:
\begin{align}
N^{\rm inj}_{\nu_{e}}(E_{\nu},r) &=\frac{4 c n_{\rm CSM}(r) m_{\rm e}}{v_{\rm sh} m_{\rm p}}\int_{m_{\rm p}}^{\infty} \mathrm{d} E_{\rm p} \frac{\sigma_{\rm pp}(E_{\rm p})}{E_{\rm p}} N_{\rm p}(E_{\rm p},r) F_{\nu_{e}}\left(\frac{E_{\nu}}{E_{\rm p}} \right) \nonumber\\
 N^{\rm inj}_{\nu_{\mu}}(E_{\nu},r) &=\frac{4 c n_{\rm CSM}(r) m_{\rm e}}{v_{\rm sh} m_{\rm p}}\int_{m_{\rm p}}^{\infty} \mathrm{d} E_{\rm p} \frac{\sigma_{\rm pp}(E_{\rm p})}{E_{\rm p}} N_{\rm p}(E_{\rm p},r) \nonumber \\ & \hspace{6cm} \times \left[F_{\nu_{\mu}}^{(1)}\left(\frac{E_{\nu}}{E_{\rm p}} \right)+F_{\nu_{\mu}}^{(2)}\left(\frac{E_{\nu}}{E_{\rm p}} \right)\right]\ .
\label{nu_source2}
\end{align}
Here, $F_{\nu_{e}}$, $F_{\nu_{\mu}}^{(1)}$, $F_{\nu_{\mu}}^{(2)}$ and $f_{\nu_{e}}$, $f_{\nu_{\mu}}^{(1)}$, $f_{\nu_{\mu}}^{(2)}$ 
are the neutrino distributions from  pion decay valid for $E_p\geq0.1$ TeV and  $E_p<0.1$ TeV, respectively~\cite{Kelner:2006tc}. The energy dependent cross-section $\sigma_{\rm pp}(E_{\rm p})$ is taken from Ref.~\cite{Kelner:2006tc}.  The neutrino flux of flavour $f$ at source is given by,
\begin{align}
  & E_{\nu}^2 \phi^{\rm s}_{\nu_{\rm f}}(E_{\nu}) =\int_{r_{\rm in}}^{r_{\rm max}}\frac{\mathrm{d}r}{v_{\rm sh}} \frac{E_{\nu}^2  N_{\nu_{\rm f}}(E_{\nu},r)}{ m_{\rm e} c^2  t_{\rm esc}}.
\label{eq:totNuSource}
\end{align}
The all flavour flux at source is $E_{\nu}^2\phi_{\nu}^{\rm s}=\sum_{\rm f}E_{\nu}^2 \phi^{\rm s}_{\nu_{\rm f}}(E_{\nu})$.  During propagation to Earth, the flavour combination may change due to flavour mixing~\cite{Anchordoqui:2013dnh}, leading to the following  flavour composition at Earth:  $\nu_{\rm e}:\nu_{\mu}:\nu_{\tau}=1:1:1$. 
\begin{equation}
      E_{\nu,\rm obs}^2 \Phi_{\nu}(E_{\nu,\rm obs}) = \frac{\sum_{\rm f}E_{\nu}^2 \phi^{\rm s}_{\nu_{\rm f}}(E_{\nu})}{4 \pi d_{\rm L}^2},
\end{equation}
where $ E_{\nu} = (1+z) E_{\nu,\rm{obs}} $ and $z$ is the cosmological redshift. 

Note that the neutrino (gamma-ray) production from $pp$ interactions ceases to be significant as the shock radius reaches the maximum radius, $r_{\rm max}$. This is determined by $\min[r_{\rm dec}, r_{\rm out}]$.  The deceleration radius, $r_{\rm dec}$, of the SN shock corresponds to the radius when the ejecta mass, $M_{\rm ej}$, sweeps up the same amount of  CSM mass and is defined as $r_{\rm dec}=M_{\rm ej} v_{\rm w}/\dot{M}_{\rm W}= r_{\rm in} + {M_{\rm ej}}/{(4 \pi m n_{\rm in,CSM} r_{\rm in}^2)}$. The region $r<r_{\rm dec}$ is known as the free expansion  phase of the CSM. Particle acceleration and production are dominant in this phase. Once the shock reaches the maximum radius i.e., $r_{\rm max}$, the interaction of the shock with the ISM begins, which is beyond the scope of this work\footnote{In the case of old SNR ($~1$ kyr), the flux due to the Sedov-Taylor phase ($r>r_{\rm dec}$) decreases by several (more than $3$--$4$) orders of magnitude  due to the very low target (CSM or ISM) density. Thus, the total flux from the Sedov-Taylor phase will be negligible compared to the early YSN phase. }. 

Most SNe show signs of interaction within a few years of their evolution with rarer cases showing interaction on timescales of $\mathcal{O}(10)$~years, see e.g.~Refs.~\cite{Smith:2016rwo, Pooley:2019ydb, Chandra:2020bzz,Balasubramanian:2021rjv,Szalai:2021sii, 2022arXiv220307388D}.   Since the maximum time, $t_{\rm max}$ (corresponding to $r_{\rm max}$) depends on the extent of the CSM,  YSNe with extended CSM could have  $t_{\rm max} \sim \mathcal{O}(10)$~years.
However, this extended CSM is less dense than the CSM of $t_{\rm max} \simeq 1$~year. Therefore, the neutrino and gamma-ray fluxes for this extended CSM are similar to that of  YSNe with CSM of 1 year (results not shown here).  Therefore, we take $r_{\rm max}= r_{\rm out}$ (corresponds to $t_{\rm max}= 1$ year) throughout this work.

Different classes of YSNe produce different fluxes of secondaries (gamma-rays and neutrinos) due to wide differences in these input parameters.   In the following, we discuss the different Types of YSNe and their impact on the secondary production.

\section{Young supernova types}
\label{sec:DOSNR}
Supernovae are classified as Type-I and Type-II based on the presence of H lines in the observed spectra. Type-I SNe do not show H lines, whereas Type-II do. Supernovae are further classified into different sub-Types of Type-I (Ia, Ib and Ic) and Type-II (IIn, II-P, IIb, II-L) SNe depending on different observed characteristics. Type Ia SNe show Silicon II (Si II) absorption lines in their spectra. Presence of He lines in Type Ib SNe differentiates them from Type Ic. Type IIn and IIb exhibit narrow and broad H lines respectively. The spectra of Type II-P SNe show a plateau whereas Type II-L SNe show linearly declining spectra (for details on SN classification, see 
e.g.~Refs.~\cite{Gal-Yam:2016yms,Turatto:2003np}).  Except for SNe Ia, all other SN classes  are powered by the  collapse of the stellar core.  

The diverse classes of YSNe differ in parameters like mass loss rate, wind velocity, size of the CSM, and efficiency of shock energy transfer to protons. Below we report about  these characteristic parameters as from electromagnetic observations for different Types of YSNe.

\begin{itemize}
\item Type IIn SNe are found to have very dense CSM as a result of heavy mass loss ($10^{-4}$--$10^1$ $ \rm \dot{M}_{\odot} yr^{-1}$) before explosion \cite{Smith:2014txa,Moriya:2014cua}. The wind velocity, $v_{\rm w}$, for such SNe can typically vary between $30$--$600~\rm km s^{-1}$ \cite{Smith:2014txa}. Their progenitors are not well known and have been found to be connected to LBV, Red Supergiant (RGB) and Yellow Hypergiant (YHG) stars (see e.g., \cite{Smith:2016dnb,Taddia:2013nga,Ransome:2021bfw,2011MNRAS.412.1639D} and references therein). These stars  have large mass loss rates ($\dot{M}_{\rm W} \sim 10^{-3}$-$10^1 ~\rm M_{\odot} yr^{-1}$)  and moderate wind velocity ($v_{\rm w}\sim 100~\rm km s^{-1}$)~\cite{Smith:2014txa,Moriya:2014cua}. 
Therefore, Type IIn SNe show a dense CSM ($n_{\rm in, CSM}\sim 10^{11}$ $\rm cm^{-3}$~\cite{Smith:2014txa}). Their shock velocity, $v_{\rm sh}$, ranges between $10^3$ and $10^4$ $\rm km\ s^{-1}$~\cite{Petropoulou:2017ymv,Ofek:2013afa,Ofek:2014fua} (see Eq.~\ref{eq:CSM_density}). The fraction of kinetic energy, $\epsilon_{\rm p}$,  that goes into  high energy protons is estimated to be  $10^{-2}$--$1.5 \times 10^{-1}$~\cite{Caprioli_2014}.
Therefore, we choose the range of $\epsilon_{\rm p}$ as  $10^{-2}$--$10^{-1}$.  The fraction of post-shock magnetic energy, $\epsilon_{\rm B}$, can be constrained from radio observations of SNe and is in the range $\epsilon_{\rm B} \sim 10^{-4}$--$10^{-2}$ \cite{Ofek:2013afa,Ofek:2013jua,Chandra:2008aq,Chevalier:1999mi}.

\item Type II-P SNe show a plateau in their light curves up to a few months ($\sim 20$--$100$~days) after the explosion \cite{Morozova:2016efp,Morozova:2016asf}. This is due to the abundance of hydrogen in the SN ejecta. 
This SN class usually arises from red supergiant (RSG) stars~\cite{Smith:2014txa,2012A&A...537A.146E,Smartt:2015sfa,Heger:2002by,Wagle:2019aqr} and they are found to have large CSM density. Eruptive mass losses of such stars just before [$\mathcal{O}(1)$~year] the SN explosion  are believed to be the reason of their large CSM densities~\cite{Morozova:2016efp,Nakaoka:2018nfl}. The typical mass loss rate during this phase has been  reported to be $\mathcal{O}(10^{-3})~\rm \dot{M}_{\odot}$yr$^{-1}$~\cite{Nakaoka:2018nfl,Yaron:2017umb,Bullivant:2018tru} with wind velocity of $\mathcal{O}(10)~\rm km \ s^{-1}$ \cite{Margutti:2016wyh}. Larger mass loss rates [$\mathcal{O}(1) \rm \dot{M}_{\odot}yr^{-1}$] with wind velocity $100~{\rm km \ s^{-1}}$ have also been predicted, see \cite{Morozova:2016asf,Morozova:2016efp}; however, this might be an exception (see e.g., \cite{Moriya:2018eto,Jacobson-Galan:2021pki,Forster:2018mib})~\footnote{Ref.~\cite{Morozova:2016asf}  investigated early light curves of some Type II SNe.  Their results suggest the need for local CSM ($r<2 \times 10^{14}$~cm) in order to reproduce the rapid rise time and brighter emission at peak observed in some SNe. This is an exception and we therefore focus on the average mass loss over a longer period of time. 
}. Interestingly, our calculation only depends  on the ratio $\dot{M}_{\rm W}/{v_{\rm W}}$ (see Eq.~\ref{eq:CSM_density}); hence,  given the evidence of large mass loss rates, we adopt $10^{-2}$ $\rm \dot{M}_{\odot} yr^{-1}$ as our typical mass loss rate with 
 wind velocity  $\mathcal{O}(100)$~$\rm km\ s^{-1}$~\cite{Nakaoka:2018nfl,Morozova:2016asf}.  This choice is equivalent to assume a mass loss rate $10^{-3}~ \rm \dot{M}_{\odot} yr^{-1}$ with wind velocity  $10~\rm km\ s^{-1}$.

The eruptive mass loss is responsible for a dense CSM which is found to exist up to $\sim 10^{15}~\rm cm$ \cite{Yaron:2017umb}. 
The CSM density beyond this radius is smaller and 
is due to  the   stellar wind of RSG stars~\cite{Smith:2014txa}. A typical  RSG wind has a mass loss rate in the range $10^{-6}$--$10^{-5}$ $\rm \dot{M}_{\odot}$ yr$^{-1}$ with wind velocities of about $10$--$20$~$\rm km/s$~\cite{Smith:2014txa}. We distinguish these two different phases of the CSM by naming them ``eruptive'' (for the large mass loss rates) and ``normal'' (the smaller RSG winds). The shock velocity for this YSN class is 
$\mathcal{O}(10^4)$~$\rm km/s$~\cite{Bullivant:2018tru}.

\item Type II-L SNe have  linear light curves, while Type II-b show broad He lines in their spectra. Their progenitors have similar mass loss rates, which lie in the range $10^{-5}$--$10^{-4}$ $\rm \dot{M}_{\odot} {\rm yr}^{-1}$ and have wind velocity between $2\times 10^1$--$10^2$ $\rm km\ s^{-1}$~\cite{Sravan:2020tdg,Smith:2014txa,Ouchi:2017cza,Bostroem:2019ont,Das_2017,Reynolds:2019zbv,Jacobson-Galan:2021pki}.    The shock velocity for  these Types of SNe is found to be of the order of  
$10^4$ $\rm km/s$~\cite{Kamble:2015xza}. 

\item Type Ib/c SNe are rich in helium (He) and usually have lower mass loss rates ($10^{-7}$--$10^{-4}$ $\rm \dot{M}_{\odot}~yr^{-1}$) than IIn  and II-P SNe~\cite{Smith:2014txa,Gilkis:2021uht,Jung:2021pjj,YoungSupernovaExperiment:2021fur}. The wind velocity for Ib/c SNe  is very large ($v_{\rm w}\sim10^3$ $\rm km~s^{-1}$). Moreover, the inner radius of the CSM is small ($r_{\rm in} \simeq 10^{12}$~cm, corresponds to the size of the progenitor). This implies that the volume of the CSM shell in the vicinity of $r_{\rm in}$ is also small and therefore the  total number of proton targets  is quite low. Thus, the neutrino and gamma-ray emission  from such YSNe is negligible (as most of the secondaries are produced near $r_{\rm in}$). However,  late time observations of  SN 2014C (which was initially classified as Ib/c SN)  highlight   the presence of interaction of the SN ejecta with the hydrogen rich CSM at $6 \times 10^{16}$ cm~\cite{Margutti:2016wyh}. The CSM is found to  extend up to   $2.5 \times 10^{17}$~cm~\cite{Brethauer:2020bmy}; hence,  we assume $r_{\rm out}=10^{17}$ cm. This corresponds to the beginning of shock interaction at about $1.5$ years after the explosion and substantial reduction of the dense CSM interaction at about $3$ years. Thus the duration of emission is considered to be about $1.5$ years. This dense CSM is possibly created by large mass loss rate of $\mathcal{O}(1)~\rm M_{\odot}yr^{-1}$ with wind velocity in the range ($10^1-10^3~\rm km\ s^{-1}$)~\cite{Margutti:2016wyh}. However, smaller mass loss rate of $\mathcal{O}(10^{-3})~{\rm M_{\odot}yr^{-1}}$  with velocity $100~ {\rm km \ s^{-1}}$ has also been proposed~\cite{Tinyanont:2019qol}. Hence, we assume a mass loss rate of $\mathcal{O}(10^{-2})~{\rm M_{\odot}yr^{-1}}$ and wind velocity $100~{\rm km \ s^{-1}}$ to be conservative. The exact mechanism of this large mass loss is not well understood. Therefore, we assume a wind profile in order to obtain a conservative estimate of the CSM density. The deceleration radius, $r_{\rm dec}~(9\times 10^{16}~{\rm cm})$ of the shock moving forward in the CSM  is comparable to $r_{out}$. SN 2004dk and SN2019yvr are similar examples of Ib/c SN with  late time interaction after explosion~\cite{Mauerhan:2018wes,Kilpatrick:2021hup}. Reference \cite{Margutti:2016wyh} investigated a  sample of $183$~Ib/c SNe and found that $10\%$ of the sample has late time interaction like SN2014C, which is about $2.6\%$ of the total core-collapse SNe. Hence, we include  late time interaction of  Ib/c SNe in our analysis. 
We choose the SN 2014C as a representative of the class Type Ib/c late time (LT) SN
\cite{Tinyanont:2019qol,Vargas:2021bgb,Margutti:2016wyh,Murase:2018okz}. 

\end{itemize}

In Table \ref{tab:parameters}, we list the typical values adopted for each of our modelling parameters for all SN Types introduced above.  While we chose to be conservative in our choice of the characteristic parameters, we also consider a range of variability for the most uncertain model parameters and  the latter defines a band in the final results (see Sec.~\ref{sec:diffuse} and \ref{tab:uncertainty}).

\begin{table}[t]
\caption{Characteristic parameters for different  YSN Types.  We consider two different phases of the CSM of II-P SNe namely, eruptive (created by the large mass loss occurring a few years prior to explosion) and normal  (due to  the usual RSG wind during stellar evolution). 
}
\centering
\begin{tabular}{|c||c|c|c|c|c|c|}
\cline{1-7}
 {\bf Parameters} &  {\bf  IIn} &  \multicolumn{2}{c|}{\bf II-P}   &  {\bf IIb/II-L} &  {\bf Ib/c} & {\bf Ib/c (LT)} \\
\cline{3-4} & ($1$~yr) & {\bf Eruptive} & {\bf Normal} &($1$~yr) & ($1$~yr)& (1.5 yrs) \\
 & & ($\le12$) & ($>12$ days) & & &\\
\hline
\hline
$n_{\rm in,CSM}$ ($\rm cm^{-3}$) & $ 10^{11}$ & $10^{12}$ & $ 10^{9}$ & $6 \times 10^{11}$ & $2.4 \times 10^{12}$ & $2\times 10^6$ \\
\hline
$\dot{\rm M}_W (\rm M_{\odot}~yr^{-1}$ ) &  $10^{-1}$ &  $10^{-2}$ & $2 \times 10^{-6}$ & $3 \times 10^{-5}$ & $10^{-5}$ & $2\times10^{-2}$ \\
\hline
$v_{\rm w}$ ($\rm km s^{-1}$) & $10^2$ & $10^2$ & $1.5\times 10^1$ & $3 \times 10^1$ & $10^3$ & $10^2$\\
\hline
$r_{\rm out}$ ($\rm cm$) &  $2 \times 10^{16}$ & $ 10^{15}$ & $3 \times 10^{16}$ & $3 \times 10^{16}$ & $6 \times 10^{16}$ & $10^{17}$\\
\hline
$v_{\rm sh}$ ($\rm km s^{-1}$) & $7 \times 10^{3}$ & $  10^{4}$ & $ 10^{4}$ & $ 10^{4}$ & $2\times 10^4$ & $10^4$ \\
\hline
$r_{\rm in}$ ($\rm cm$) & $4 \times 10^{14}$ & $6 \times 10^{13}$ & $6 \times 10^{13}$ & $6 \times 10^{12}$ & $3 \times 10^{11}$ & $5\times10^{16}$\\
\hline
$\epsilon_{\rm p}$ & $ 10^{-2}$ & $10^{-1}$ & $10^{-1}$ & $10^{-1}$ & $ 10^{-1}$ & $5\times 10^{-2}$\\
\hline
$\epsilon_{\rm B}$ & $ 10^{-2}$ & $ 10^{-2}$ & $10^{-2}$& $10^{-2}$ & $ 10^{-2}$ & $10^{-2}$\\
\hline
\end{tabular}
\label{tab:parameters}
\end{table}

\section{Dependence of the  gamma-ray and neutrino emission on the  young supernova type}
\label{sec:point sources}
The YSN emission properties largely vary across different SN Types. In this Section, we first characterize the gamma-ray and neutrino emission for our benchmark YSN IIn and then compare the particle production across different YSN classes. Throughout this Section, the emission of  secondaries is computed for $1$ year for all YSN Types, except for Ib/c (LT) YSNe for which we consider an emission period of $1.5$ years. 

\subsection{Characteristic gamma-ray and neutrino emission}
\begin{figure}
    \centering
      \vbox{\includegraphics[width=0.48\textwidth]{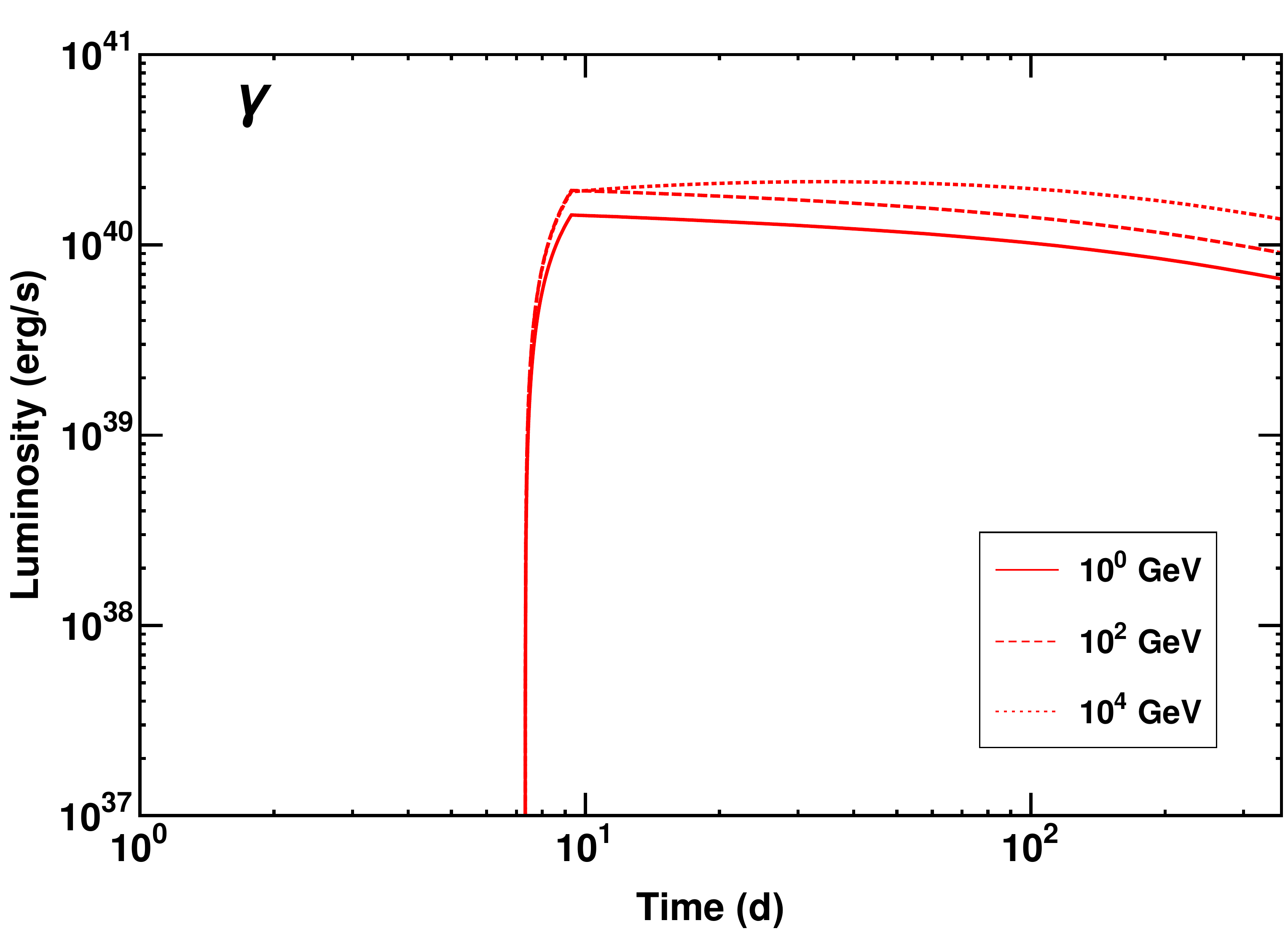}
      \includegraphics[width=0.48\textwidth]{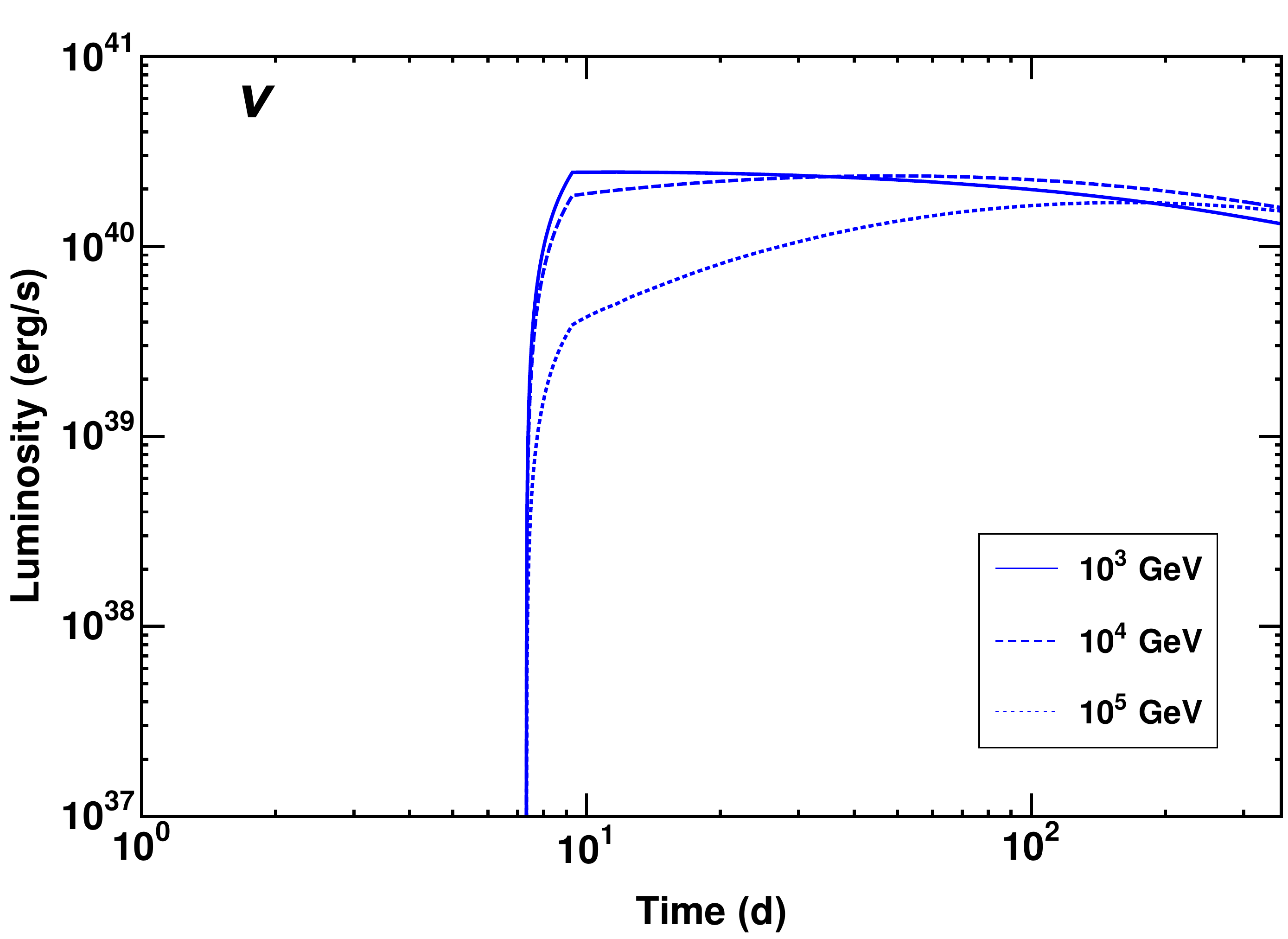}}
       \vbox{\includegraphics[width=0.49\textwidth]{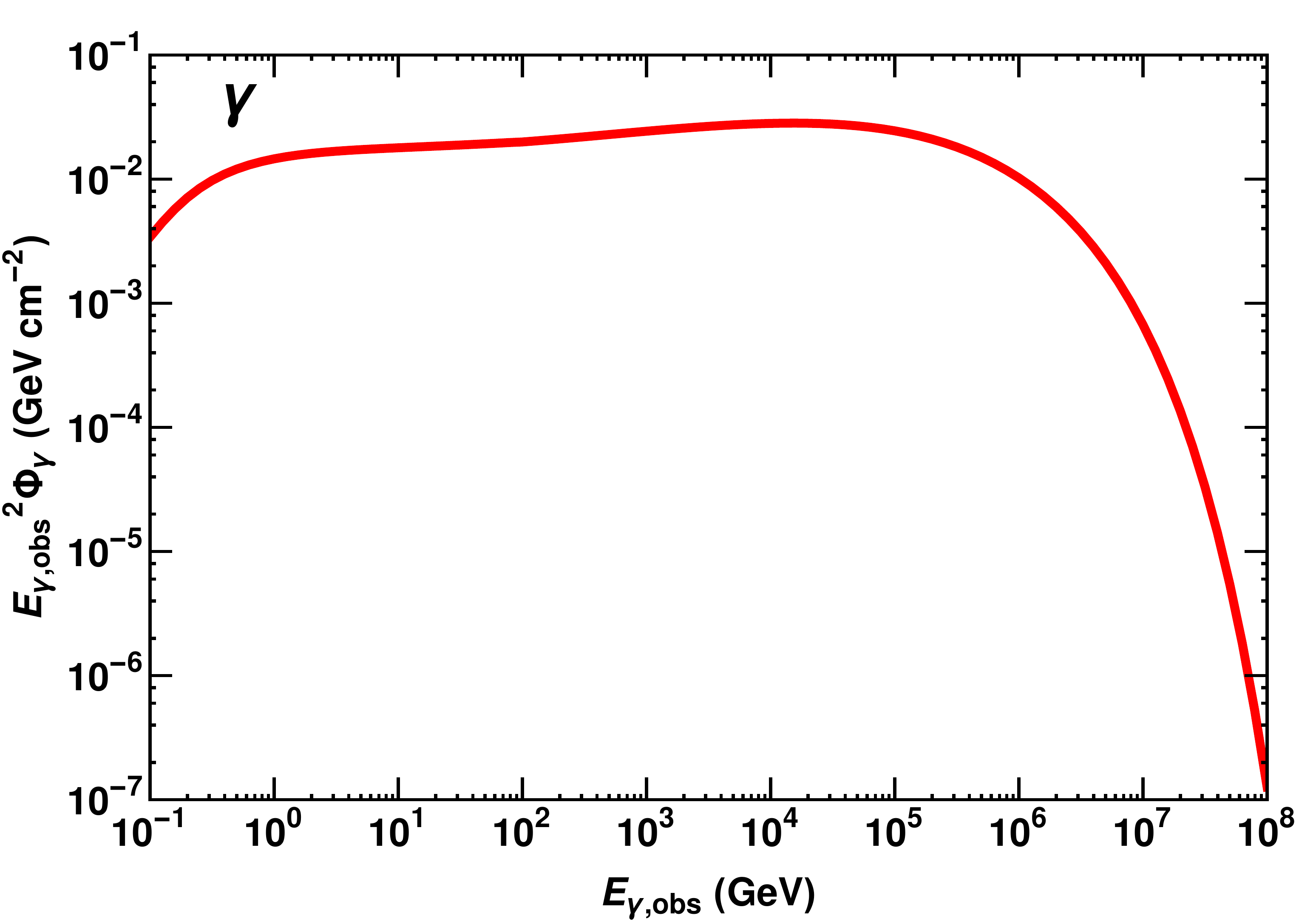}
      \includegraphics[width=0.49\textwidth]{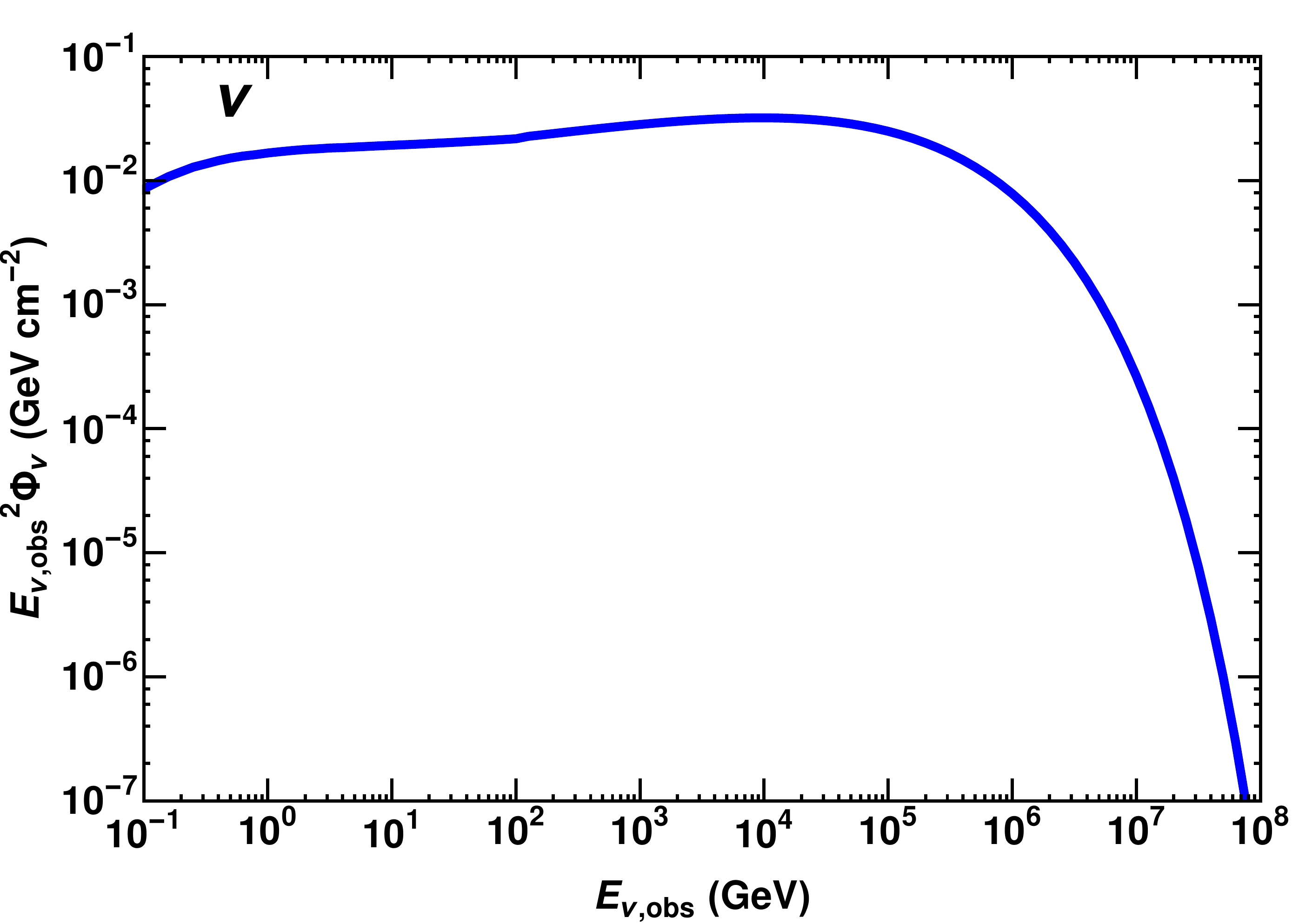}}
    \caption{{\it Top panels:} Luminosity of gamma-rays (left) and all-flavour neutrinos (right)  of our benchmark YSN IIn (see Table \ref{tab:parameters}) as a function of time. 
    The temporal evolution at the source are shown for three representative energies ($E_\gamma = 10^0, 10^2$, and $10^4$~GeV for gamma-rays and $E_\nu = 10^3, 10^4$, and $10^5$~GeV for neutrinos). 
    {\it{Bottom panels:}} Corresponding gamma-ray (left) and all-flavour neutrino (right) fluences at Earth as  functions of the observed particle energy. The distance of our YSN IIn from Earth is chosen to be  $10$~Mpc. 
    Both fluences for neutrinos and gamma-rays are time integrated over the first year of the SN evolution. Gamma-rays are shown without any attenuation due to energy losses (see discussion of Fig.~\ref{source gamma} for details).  Therefore, the spectra of both species from $pp$ collision follow a similar behaviour of growing with particle's energy and falling rapidly above $10^5$ GeV. 
    } 
    \label{fig:point_source_IIn}
\end{figure}

As a primary example to illustrate the YSN neutrino and gamma-ray emission we consider Type IIn SNe (see Sec.~\ref{sec:SNmodel} and Table~\ref{tab:parameters})  as they have the highest mass loss rate in  the SN sample presented in Table \ref{tab:parameters}. Unless otherwise specified, we assume that our benchmark YSN is located at $10$~Mpc from Earth. Most SNe are detected between $1$ and $100$~Mpc, with the majority being discovered at $\mathcal{O}(90)$~Mpc, see e.g.~Fig.~2 of Ref.~\cite{Richardson:2014gqa}. Hence, our choice of the typical SN distance is somewhat optimistic, yet compatible with observations.

\begin{figure}[t]
\centering
\includegraphics[scale=0.4]{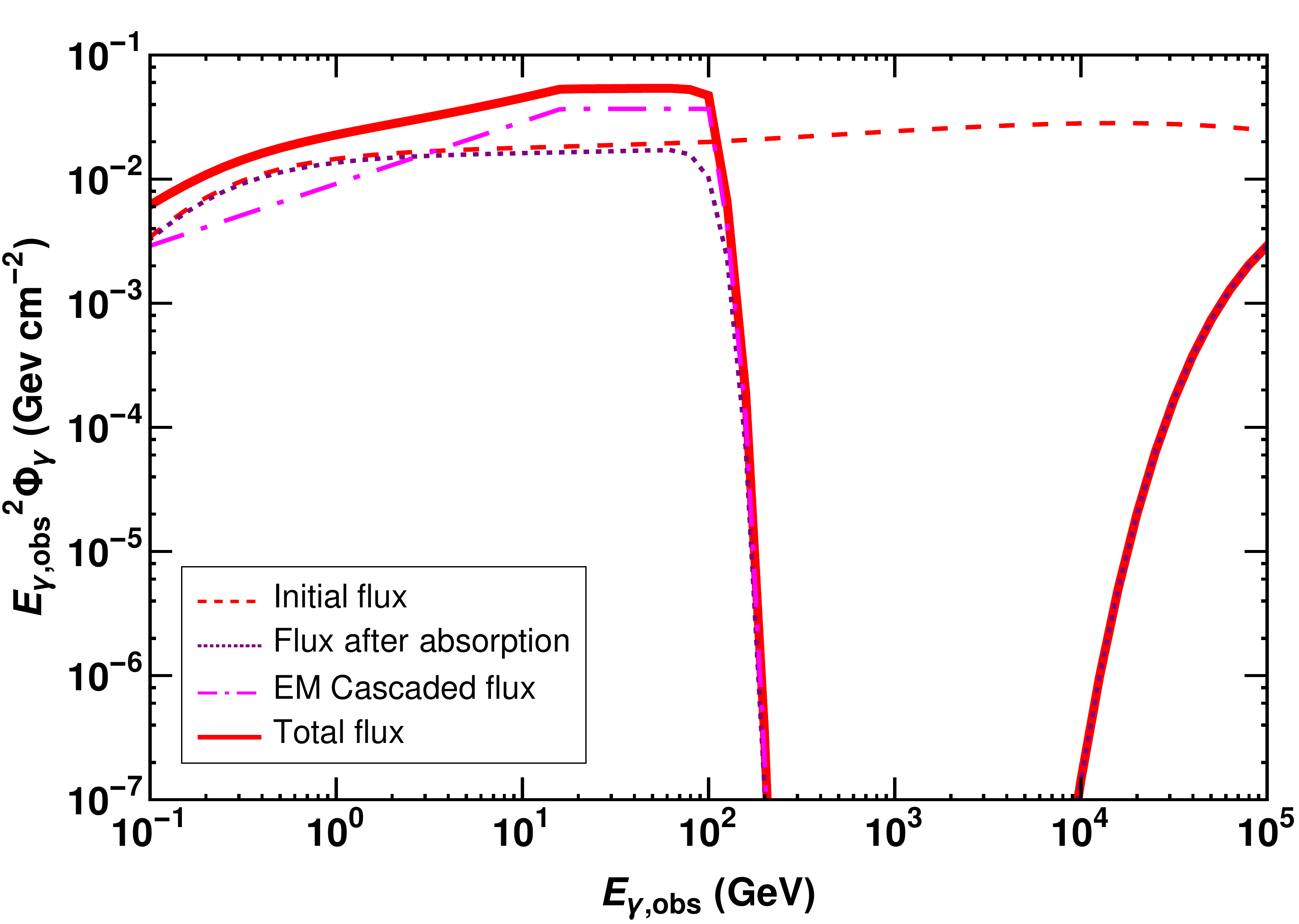}
\caption{Gamma-ray fluence at Earth after including the relevant absorption processes and EM cascades at the source as a function of the gamma-ray energy for our benchmark YSN IIn (see Table~\ref{tab:parameters}) at a distance of $10$~Mpc.
The red-dashed curve represents the fluence obtained via  $pp$ collisions. The purple-dotted curve is the fluence after absorption due to $\gamma\gamma$ pair production loss and Bethe-Heitler loss in the CSM. The gamma-ray fluence due to EM cascades is shown by the dot-dashed magenta curve. The thick red solid curve represents total flux at the source after absorption and EM cascades.  The gamma-ray fluence between $10^2$ to $10^4$~GeV is heavily attenuated due to the absorption processes occurring in the source. The final gamma-ray flux (thick red curve) is a combination of the flux surviving to absorption and the EM cascaded flux.
}
\label{source gamma}
\end{figure}

The top and bottom left (right) panels of Fig.~\ref{fig:point_source_IIn} show the  gamma-ray (neutrino) luminosity and fluence. 
The top left panel shows the gamma-ray light curve  at $E_{\gamma}=$1, $10^2$, $10^4$~GeV, while the  gamma-ray fluence produced by $pp$ collisions in IIn YSNe during its first year is displayed in the bottom left panel. The fluence in the bottom left panel of Fig.~\ref{fig:point_source_IIn}  does not include any attenuation at the source. Gamma-rays above $100$~GeV can interact with  low energy photons  and  ambient matter in the source and may get lost. This corresponds to $\tau_{\gamma\gamma}$ and $\tau_{\rm BH}$, respectively, in 
Eq.~\ref{eq:gamma_at_earth} and are discussed below (see discussion of Fig.~\ref{source gamma}). \

The top panel on the right shows the neutrino light curve (blue) for  three different energies ($E_{\nu}=10^3$, $10^4$, $10^5$ GeV) for the first year of the YSN. The light curves  start around $7$~days after the explosion. This delay corresponds to the inner radius, $r_{\rm in}$ (i.e., when the interaction between the shock wave accelerated protons and the CSM protons has begun). The  inner radius, $r_{\rm in}$, also depends on the class properties of each SN Type.  
The  plot in the bottom right panel shows the energy spectra of the YSN emitted neutrinos for the first year. The neutrino flux (top right plot) falls rapidly above energy $E_{\nu}\sim 10^5$ GeV. This is due to the maximum energy of protons $E_{\rm p,max}$, which determines the  exponential cut-off of  the proton spectra. This clearly suggests that the SN properties driving  $E_{\rm p,max}$ will influence the cut-off, and the latter would be different for different YSN Types. Note that these interpretations also hold for the gamma-rays in the left panel.

The fluence  in the bottom left panel of Fig.~\ref{fig:point_source_IIn} will be affected by absorption processes. Fig.~\ref{source gamma} shows the effect of absorption and EM cascades on the gamma-rays produced via $pp$ interactions for Type IIn YSNe for a thermal energy distribution of photons with  average energy    $\epsilon_{\rm av} = 1$~eV and SN peak luminosity  $L_{\rm SN,pk} = 10^{41}$ $\rm erg/s$ respectively \cite{Gaisser:1996qe,Petropoulou:2016zar}. 
We show gamma-rays without any absorption (red-dashed  curve), gamma-rays with absorption due to photon-photon pair production and BH pair production (purple-dotted curve), and the EM cascade (magenta dot-dashed curve).   It is interesting to note that the initial flux above $200$~GeV mostly disappears except  a little tail at higher energies due to  photon-photon pair production losses.  This is due to the fact that the photon-photon cross section $\sigma_{\gamma\gamma}$ falls rapidly at higher energies and the thermal photons, $n_{\rm ph}(\epsilon)$, have a black-body spectrum; the product of both gives rise to a optical depth, $\tau_{\gamma\gamma}$ creating a well-shaped attenuation factor $e^{-\tau_{\gamma\gamma}}$. The cascaded gamma-ray flux also falls sharply at $200$ GeV because the photon-photon pair loss factor $e^{-\tau_{\gamma\gamma}}$ hardly allows for anything to survive above $200$ GeV. 

Bethe-Heitler losses generally  have a tiny effect on the gamma-ray energy distribution.
For Type IIn SNe, the amount of attenuation is the small gap between the  red-dashed  and the purple dotted curve below $200$ GeV.  Thus the final gamma-ray flux (thick red curve) is a combination of the flux surviving to absorption (purple dotted) and EM cascaded flux (magenta dot-dashed). The EM cascades are responsible for a boost in the gamma-ray spectra below 
$200$~GeV. Propagation to Earth is not important at $10$~Mpc but important for large distances ($\sim1~\rm Gpc$) as losses due to EBL will take place, for example for the diffuse flux (see Sec.~\ref{sec:diffuse}).

\subsection{Dependence on the young supernova type }
\label{sec:DOYSNe}

Because of the intrinsic  differences among the properties of the  YSN Types introduced in Sec.~\ref{sec:DOSNR}, we should expect a wide variation in the emission of  secondary particles. Below we discuss a few examples of these variations  for the secondary gamma-rays and neutrinos  for a  source at $10$~Mpc  and $\alpha_{\rm p}=2.0$.

\begin{figure}[t]
    \centering
     \vbox{\includegraphics[width=0.49\textwidth]{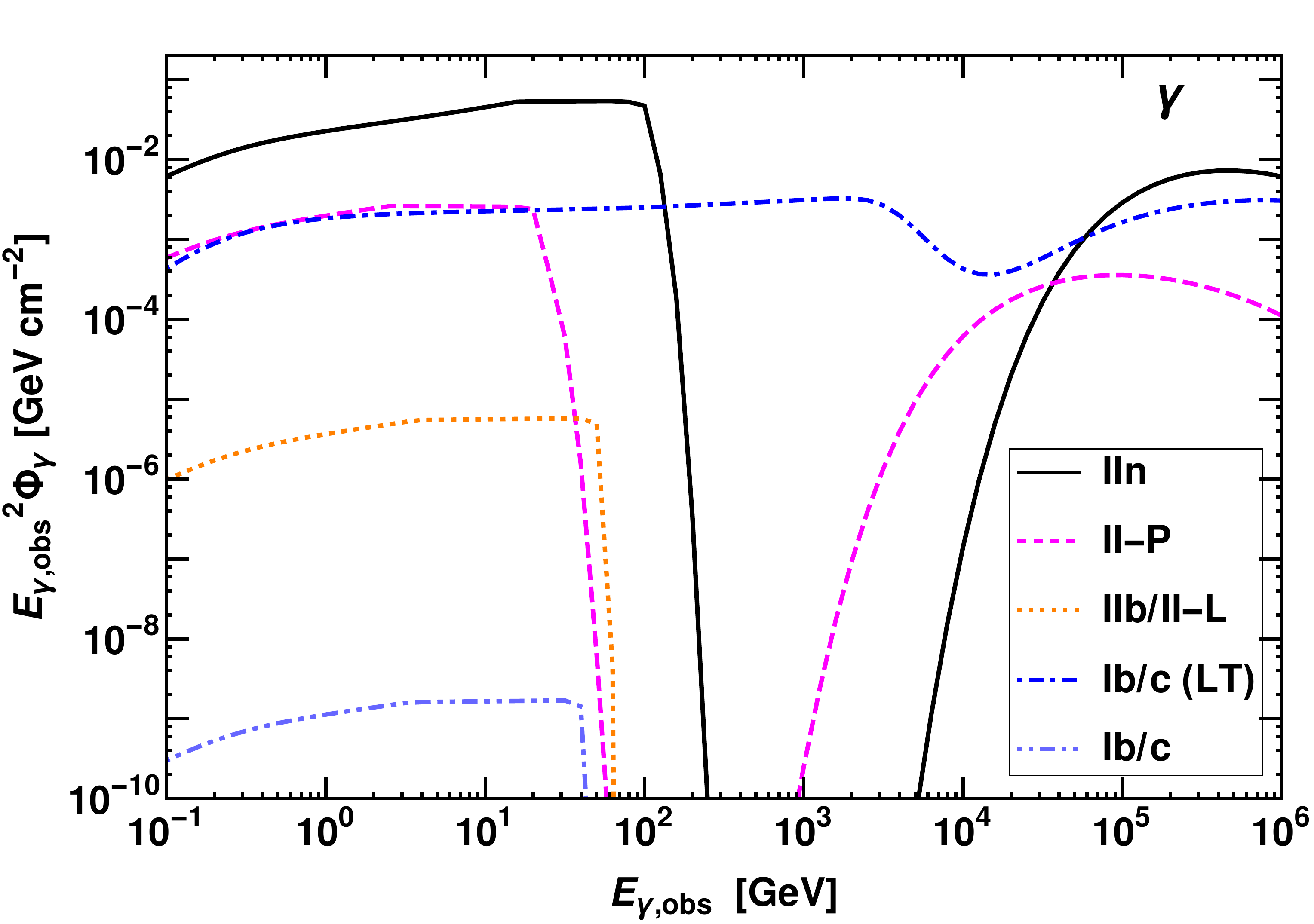}
      \includegraphics[width=0.49\textwidth]{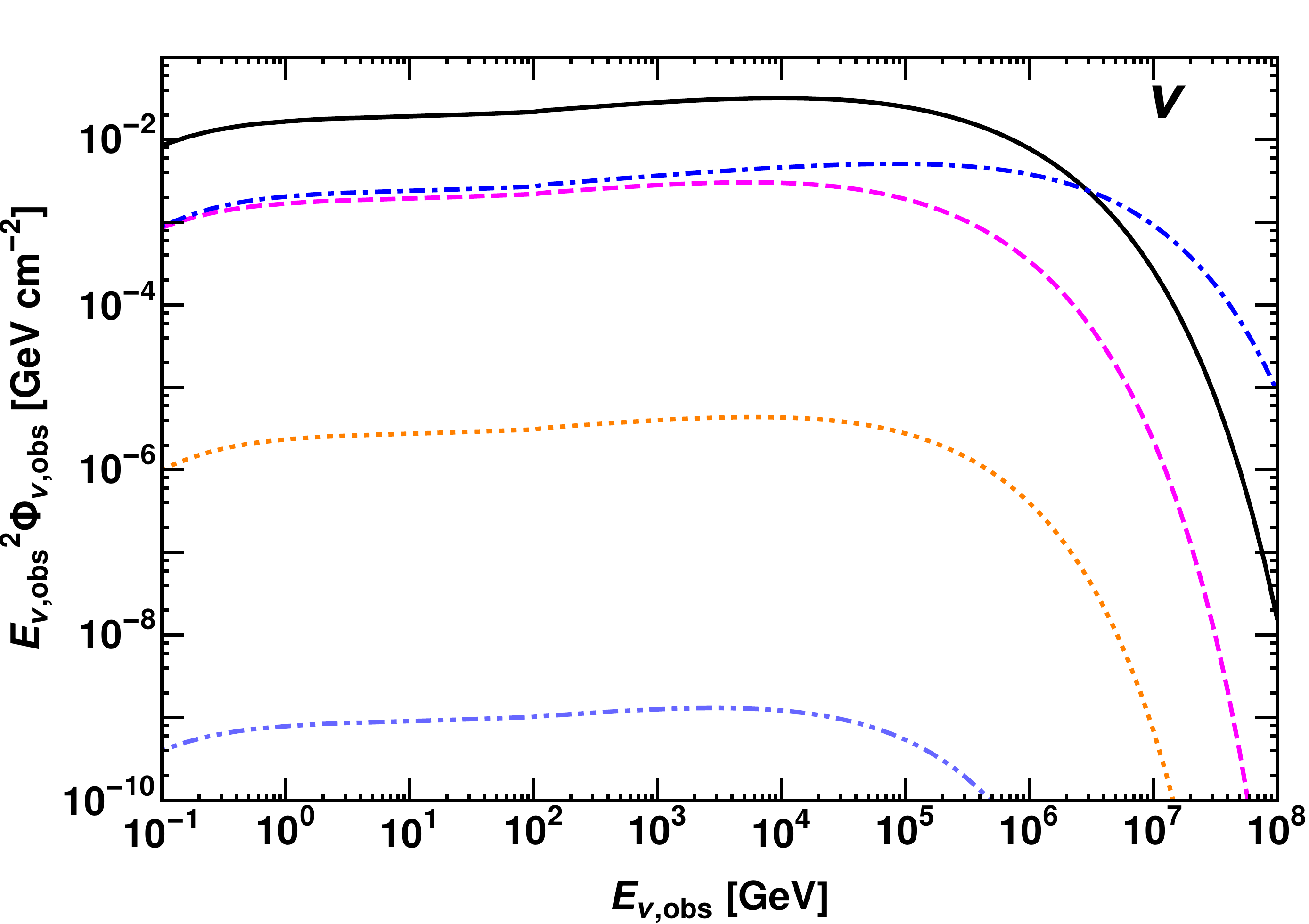}
      }
    \caption{Gamma-ray (left) and all-flavour neutrino (right) fluences integrated for a year at Earth  from different classes of YSN, as listed in Table \ref{tab:parameters}, for a source  located at $10$~Mpc. 
    For both gamma-rays and neutrinos, Type IIn YSNe have the largest fluence, followed by  II-P and  Ib/c (LT) YSNe.  
    }
    \label{fig:point_source}
\end{figure}

We  show the  gamma-ray fluences produced by YSNe in the top left panel of Fig.~\ref{fig:point_source} for the benchmark parameters shown in Table \ref{tab:parameters}. The fluxes are integrated over a year.  Type IIn YSNe produce the largest gamma-ray fluence. The other Types of YSNe create smaller fluences and the fluence of Type Ib/c YSNe is extremely small relative to the one seen from the Type IIn YSNe.
The sharp fall and rise of these fluences at higher energies (above $10$~GeV) is due to the fact that gamma-rays suffer losses due to $\gamma\gamma$ pair production, as discussed in Sec.~\ref{subsec:source abs}. The fall and rise of the fluences occur at different energies according to the  YSN class. The amount of losses are also different for different YSNe. This is due to the variation of the parameters, i.e.~average energy ($\epsilon_{\rm av}$) and peak luminosity ($L_{\rm SN,pk}$) of thermal photons expected from different Types of YSNe.    The parameter $\epsilon_{\rm av}$ is assumed to be: $1$~eV for IIn YSNe, $4.5$~eV for II-P YSNe, $1$~eV for IIb/II-L and Ib/c YSNe, and $0.05$~eV for Ib/c (LT) YSNe.  The peak luminosities  are assumed to be $10^{41}$~erg/s for IIn SNe, $\sim 5 \times 10^{40}$~erg/s for II-P and Ib/c (LT) SNe, and $10^{42}$~erg/s for IIb/II-L and Ib/c SNe~\cite{Petropoulou:2016zar,Yaron:2017umb,Miller:2008jy,Kamble:2015xza,Murase:2018okz,Immler:2007mk,Singh:2021qcx}. The fluence of Type Ib/c (LT) YSN falls at higher energy than that of other SN Types due to its small average energy of thermal photons. The average energy is small because of the lower temperature of  thermal photons as the CSM is located very far away (see Table \ref{tab:parameters}) from the stellar envelope.
The large distance to the CSM also affects the density of the thermal photons that results in less absorption of gamma-rays for Type Ib/c (LT) YSNe.   Type IIn and II-P YSNe have similar losses as the peak luminosities of thermal photons are quite similar. Heavy losses of gamma-rays are seen for Type IIb/II-L and Ib/c YSNe because their thermal photon luminosity is quite large which implies large thermal photon density.

The  right panel of Fig.~\ref{fig:point_source} shows the corresponding neutrino fluence integrated over a year  for the different YSN Types. 
Due to the dense CSM,  Type IIn YSNe show the largest fluence falling rapidly around $E_{\nu}\sim 10^6$ GeV.  The fluence of Type Ib/c (LT) YSNe 
starts to dominate above $E_{\nu}\sim 3 \times 10^6$~GeV.  The reason is that the maximum proton energy ($E_{\rm p, max}$)  strongly depends  on the shock velocity and shock radius. The shock velocity and shock radius for Ib/c (LT) YSNe are larger than those of Type IIn YSNe,  resulting in greater $E_{\rm p, max}$. The fluence of Type II-P\footnote{ The fluences of Type II-P (gamma-rays and neutrinos) are dominated by the emission from the eruptive phase (see Table~\ref{tab:parameters}), the contribution from the normal phase is negligible.} and Ib/c (LT) YSNe are comparable at lower energies due to their dense CSM. 
The fluences of Type IIb/II-L and Type Ib/c YSNe are quite small in spite of the large initial CSM density, $n_{\rm in, CSM}$.  This is due to the small size of their inner radius ($r_{\rm in}$) of the CSM; moreover, most of the secondaries are only produced in the vicinity of $r_{\rm in}$.

The estimated neutrino fluences for the various YSNe classes are comparable to the ones reported in
 Ref.~\cite{Murase:2017pfe} except for the ones of Type II-P YSNe. In this case, our YSNe-IIP fluence is larger~\footnote{Note that the large mass loss rates of II-P SNe discussed in Ref.~\cite{Morozova:2016asf} may be responsible for a fluence comparable to the one of IIn YSNe~\cite{Kheirandish-TAUP,Kheirandish:2022eox}. 
 However, such mass loss rates  might be an exception, as we discussed in Sec.~\ref{sec:DOSNR} and therefore we do not consider them. } since it relies on model parameters extrapolated from more recent  observations of II-P SNe, as discussed in Sec.~\ref{sec:DOSNR};
 specifically, the mass loss rate that we adopt is one order of magnitude larger than the one assumed in Ref. \cite{Murase:2017pfe}. On the other hand, a CSM density smaller than our benchmark case in Table \ref{tab:parameters} would result in a smaller fluence ~\cite{Murase:2017pfe}.
 In addition, the  neutrino emission from Ib/c (LT) SNe is explored for the first time in this paper.

\begin{figure}[t]
    \centering
     \vbox{
      \vspace{0.5cm}
      \includegraphics[width=0.48\textwidth]{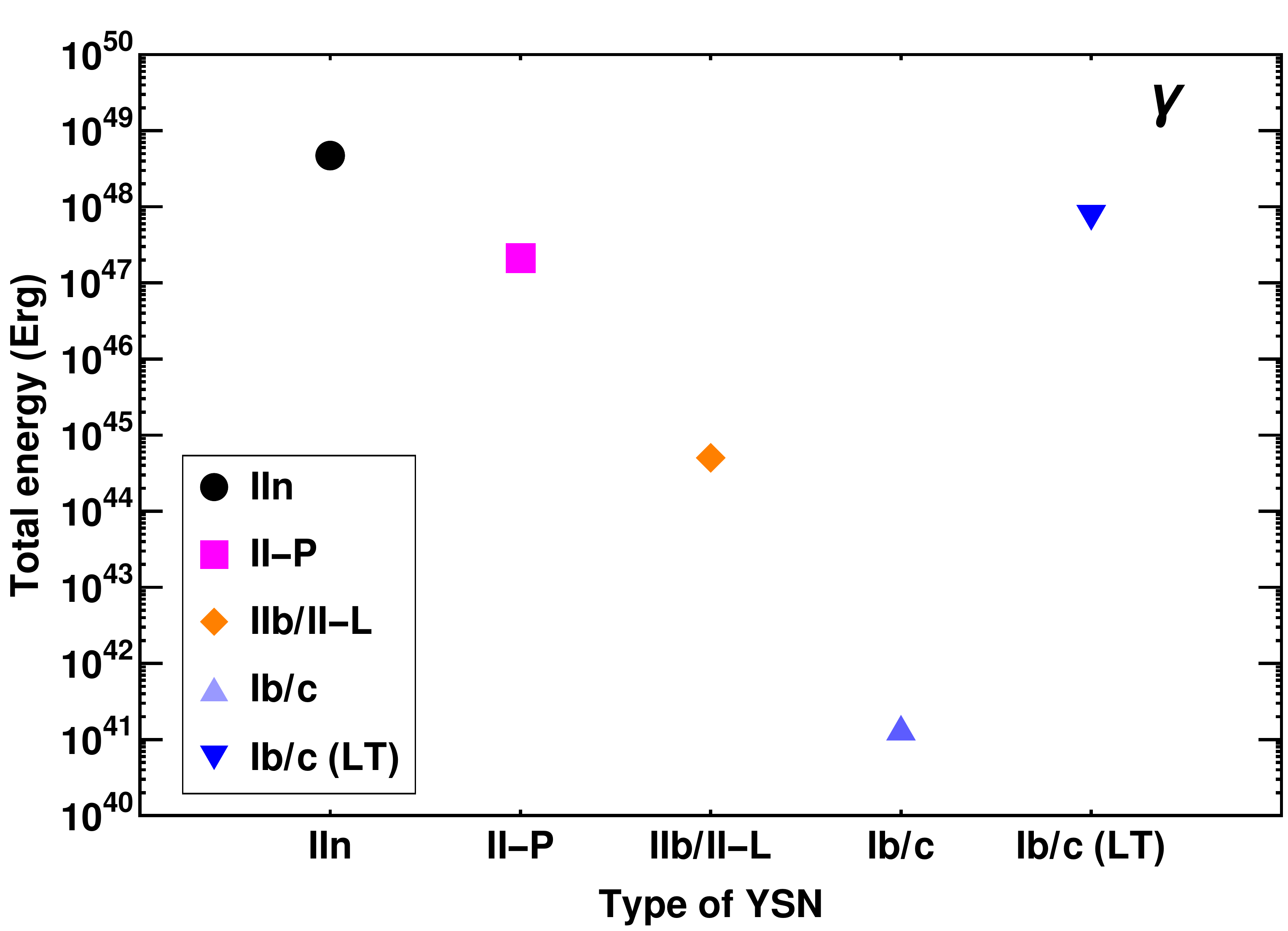}
      \includegraphics[width=0.48\textwidth]{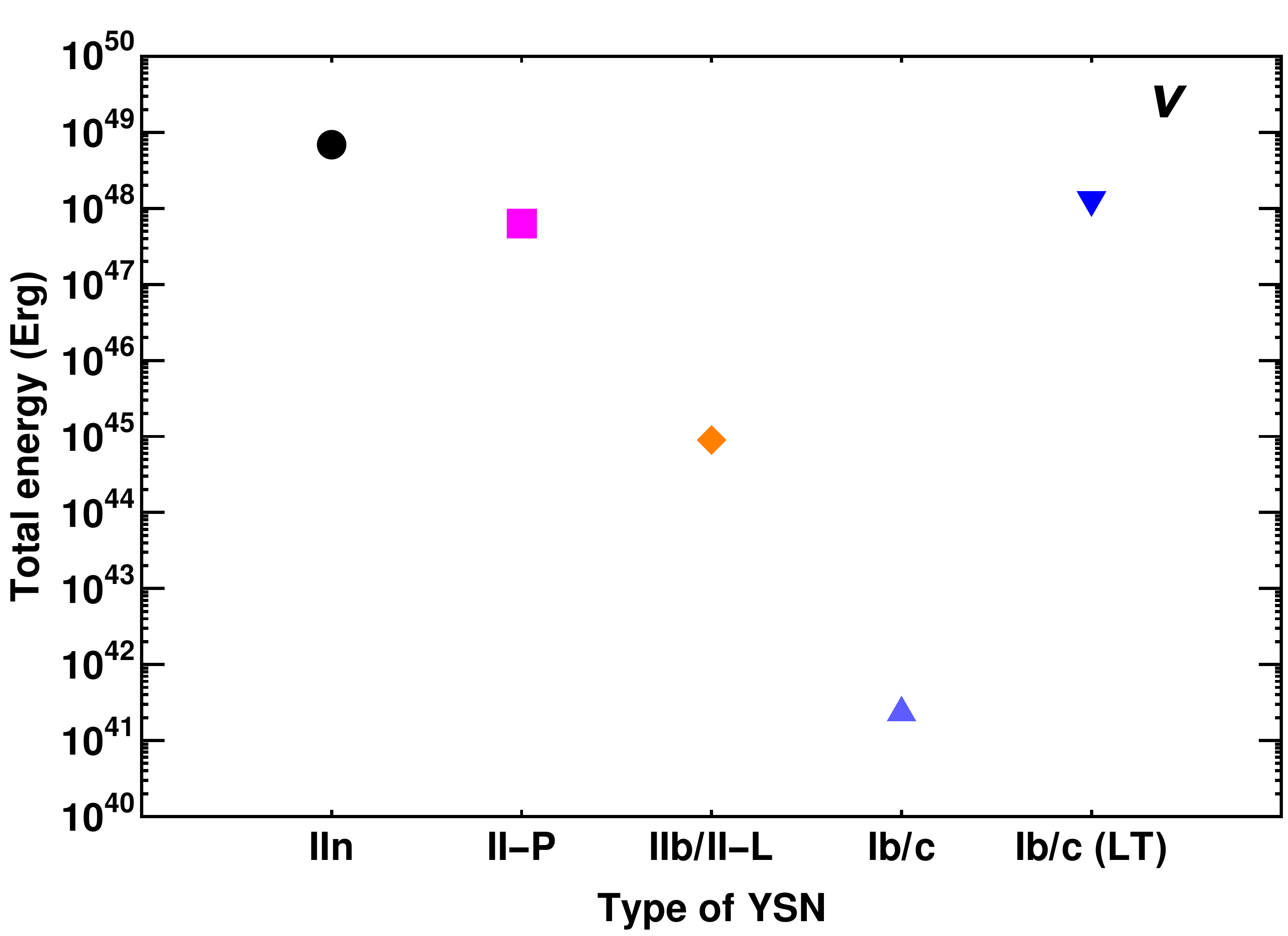}
      }
    \caption{
    Total energy emitted by each source in gamma-rays (left) and neutrinos (right). 
    The duration of emission of secondary particles is taken to be a year and integrated in the range $10^{-1}$--$10^8$ GeV. The total energy emitted in gamma-rays is generally smaller than that of neutrinos; this is due to the attenuation of gamma-rays in the source. The largest energy is emitted by Type IIn YSNe,  followed by Type Ib/c (LT)  and II-P YSNe. 
    }
    \label{fig:point_source1}
\end{figure}
Fig.~\ref{fig:point_source1}  shows the total energy emitted in  gamma-rays (left) and neutrinos (right) for each YSN Type, in order to favour a comparison among their overall energetics. We have integrated the gamma-ray and neutrino fluxes in the energy range $10^{-1}$--$10^{8}$ GeV 
for  a year. 
Both for gamma-rays and neutrinos, the largest energy is emitted by Type IIn YSNe,  followed by Type Ib/c (LT)  and II-P YSNe. 
By looking at the fluxes of neutrinos and gamma-rays in Fig. \ref{fig:point_source},  one might have the impression that  the total energy emitted in gamma-rays is smaller than that of neutrinos due to the large absorption of gamma-rays (the dips). However this is not the case, most of the attenuated high energy gamma-ray flux has reappeared at lower energies due to the EM cascades (see Fig.~\ref{source gamma} and discussion) determining an overall small  amount of loss. If there is no absorption of gamma-rays, the total energy of neutrinos is larger by a factor of about $3/2$ than that of gamma-rays. This corresponds to the product of charged to neutral pion production ratio ($2$) and the fraction of energy carried by the neutrinos in a charged pion decay ($3/4$).


\section{Diffuse  gamma-ray and neutrino backgrounds from young supernovae}
\label{sec:diffuse}
In this Section, we first introduce the procedure adopted to compute the diffuse backgrounds of high-energy particles. Then, we present our results on the gamma-ray and neutrino diffuse backgrounds from YSNe. A discussion on the model uncertainties and comparison with existing data follows.

\subsection{Diffuse  flux and its ingredients}
The differences among the different YSNe classes for what concerns the emission of secondaries give rise to interesting questions regarding the contribution of each YSN class to the diffuse emission of gamma-rays and neutrinos.  In addition to the  individual YSN fluxes, the diffuse flux of secondary particles embeds  the contribution from the redshift dependence of the various YSNe Types~\footnote{We assume a constant luminosity function based on the benchmark parameters introduced in Table~\ref{tab:parameters}, in light of the existing observational uncertainties~\cite{2011MNRAS.412.1522S,Cappellaro:2015qia}.}:
\begin{equation}
E_{\rm j,\rm obs}^2 \phi_{\rm j,\rm diff}(E_{\rm j,\rm obs})=\zeta \frac{c}{H_{0} }\int_0^{z_{\rm max}}\mathrm{d}z   \frac{R_{\rm CCSN}(z) E_{\rm j}^2\phi_{\rm j}^{\rm s}(E_{\rm j})}{\sqrt{\Omega_m(1+z)^3+\Omega_{\Lambda}}} e^{-\tau_{\rm j,EBL}(E_{\rm j},z) }\ ,
\label{eq:diff}
\end{equation}
where, ${\rm j} =\nu$ or $\gamma$, $E_{\rm j}= (1+z) E_{\rm j,obs}$ and $\tau_{\rm j,EBL}(E_{\rm j},z)$ is optical depth of EBL.
For neutrinos,  $\tau_{\nu,\rm EBL}=0$. 
For gamma-rays, the optical depth $\tau_{\gamma,\rm EBL}$ is taken from Ref.~\cite{Stecker:2016fsg}.  We adopt the $\Lambda\mathrm{CDM}$ cosmological model, with  $\Omega_m=0.31$, $\Omega_{\Lambda}=0.69$, and $H_0=68 \hspace{0.1 cm} \rm{km} \hspace{0.1cm} \rm{s}^{-1} \rm{Mpc}^{-1}$~\cite{Planck:2015fie}. 

\begin{figure}[]
    \centering
    \includegraphics[scale=0.6]{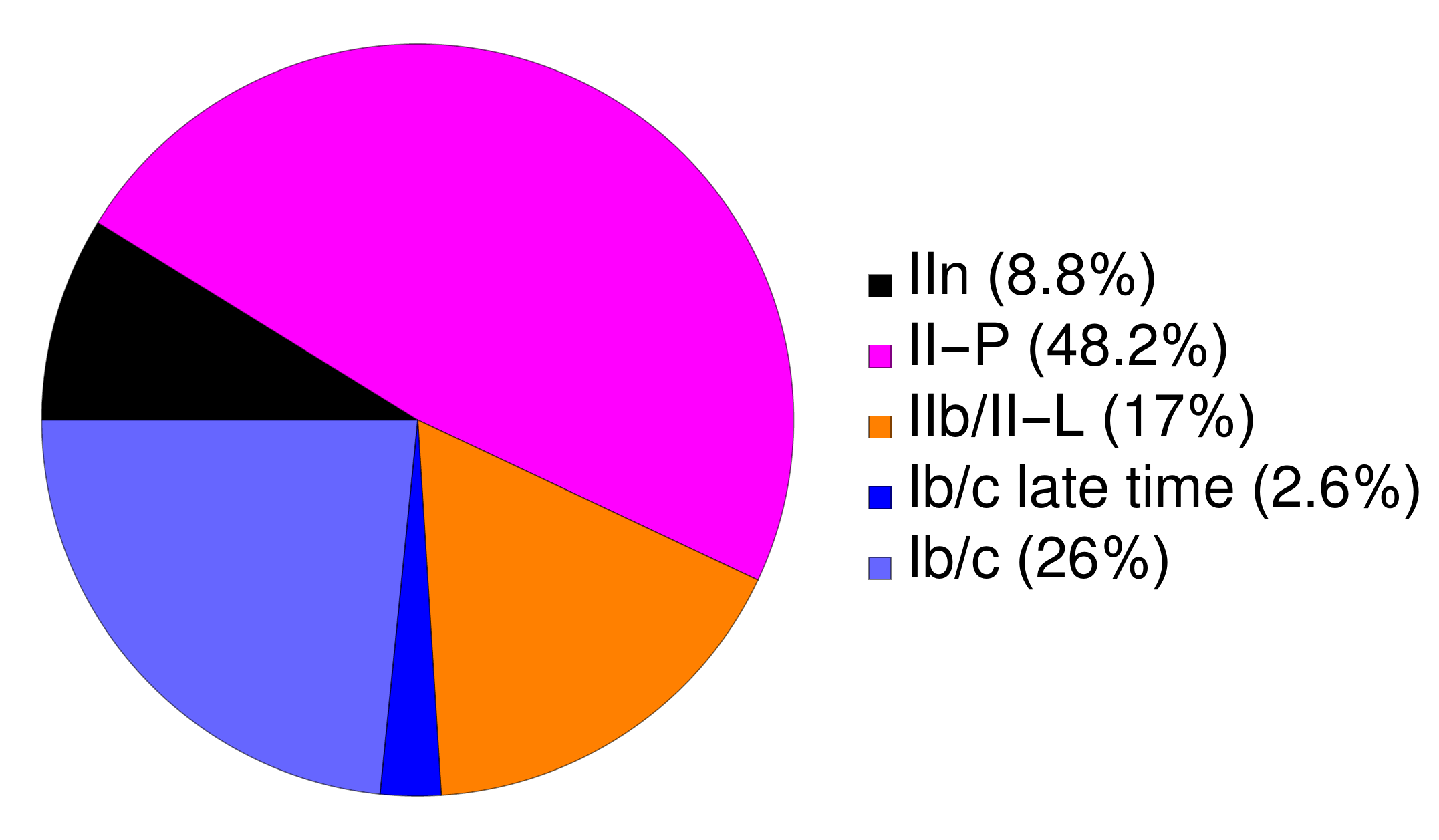}
    \caption{Local rate of   core-collapse SNe~\cite{2011MNRAS.412.1522S}. Type II-P SNe are  the most abundant ones at $z=0$. Type Ib/c and IIb/II-L SNe are also more frequent than Type IIn SNe. We assume  Type Ib/c (LT) SNe to be $10\%$ of SNe Ib/c~\cite{Margutti:2016wyh};  the total rate of SNe Ib/c (i.e., $26 \%$) includes the one of  Ib/c (LT) SNe. 
    }
    \label{fig:PiChart}
\end{figure}

The fraction of different SN Types may vary with redshift due to the change in density of stars and metallicity of the host galaxies seen at higher redshifts.  
However, unfortunately, the redshift distribution of  SNe of different  Types is quite uncertain, and limited information  is available  up to  $z=1$, for some SN Types, which is not sufficient for our purposes~\cite{Cappellaro:2015qia}. Hence, we assume that all SN Types follow the core-collapse SN rate as a function of the redshift. In addition, in order to take into account that some SN Types are more common than others, we follow Ref.~\cite{2011MNRAS.412.1522S} and assume that   the fraction of different core-collapse SN Types at $z=0$ ($\zeta$) holds at higher $z$ as well. The fraction of different SN Types at $z=0$ is shown in Fig.~\ref{fig:PiChart}.

The rate of core-collapse SNe is given by~\cite{Strolger:2015kra,2012A&A...545A..96M,Madau:2014bja}:
\begin{equation}
R_{\rm CCSN}(z)=\int_{8 M_{\odot}}^{125 M_{\odot}} \mathrm{d}M R_{\rm SN}(z,M) ,
\end{equation}
where 

\begin{equation}
R_{\rm SN}(z,M)=\frac{\eta(M)}{\int_{0.5 M_{\odot}}^{125 M_{\odot}} \mathrm{d}M M \eta(M)}R_{\rm SFR}(z) ,
\end{equation}
with $\eta(M)\propto M^{-2.35}$ being the initial mass function (following the Salpeter law)~\cite{Salpeter:1955it}. The star formation rate $R_{\rm SFR}$ is~\cite{Yuksel:2008cu}, 
\begin{equation}
R_{\rm SFR}(z)= C_0 \left[(1+z)^{p_1 k}+\left(\frac{1+z}{5000}\right)^{p_2 k}+\left(\frac{1+z}{9}\right)^{p_3 k}\right]^{1/k} ,
\end{equation}
where $k=-10$, $p_1=3.4$, $p_2=-0.3$ and $p_3=-3.5$. The constant of proportionality $C_0$ is determined by normalizing the SN rate to the local SN rate as $\int_{8 M_{\odot}}^{125 M_{\odot}} \mathrm{d}M R_{\rm SN}(0,M)=1.25 \pm 0.5 \times 10^{-4} \hspace{0.2cm} \mathrm{Mpc}^{-3}\mathrm{yr}^{-1}$~\cite{Lien:2010yb}.

\subsection{Diffuse background of gamma-rays}
\label{sec:DBGR}

The diffuse background of gamma-rays from YSNe can be computed by relying on Eq.~\ref{eq:diff}.
 Because of  $\gamma\gamma$ interactions occurring between YSN gamma-rays and the EBL,  losses similar to the ones occurring  in the source, and discussed in Sec.~\ref{subsec:source abs}, can take place. 
Substantial EBL losses  can affect the  high-energy tail of the gamma-ray spectral distribution, while gamma-rays  travel over   large distances. For example, a $300$~GeV gamma-ray photon would need to travel about $1$~Gpc to be attenuated by an amount of $1/e$~\cite{Stecker:2016fsg}.  
 
In order to compute the amount of EBL absorption, one needs to know the  redshift dependence of the EBL for different energies~\cite{Stecker:2005qs,Finke:2009xi,Stecker:2016fsg,Saldana-Lopez:2020qzx}. In this paper, we model  $\tau_{\gamma, \rm EBL}(E_{\gamma})$ (see Eq.~\ref{eq:diff}) following Ref.~\cite{Stecker:2016fsg}.

\begin{figure}[]
\centering
\includegraphics[scale=0.4]{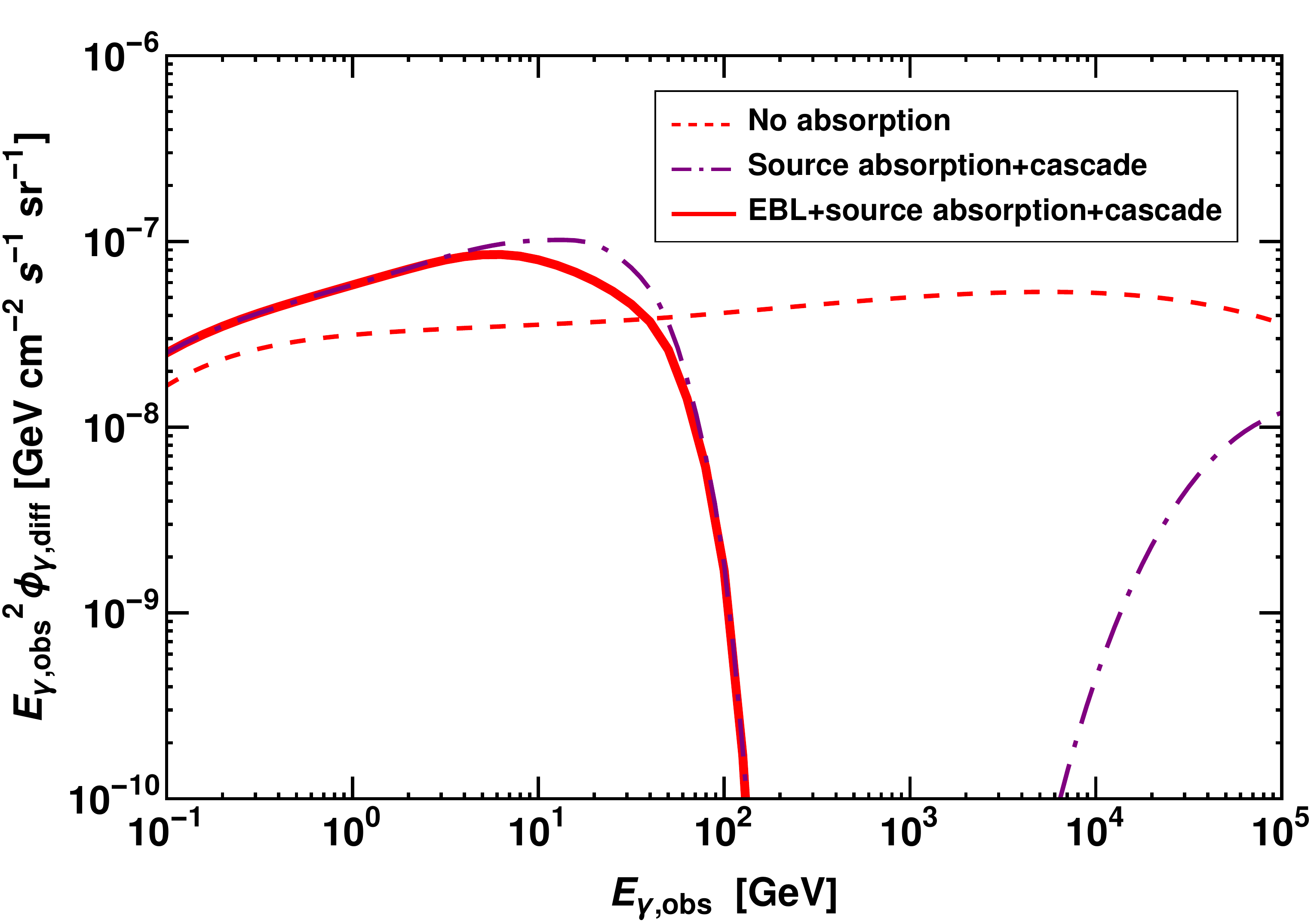}
\caption{Diffuse gamma-ray flux at Earth for our benchmark Type IIn YSN (see Table~\ref{tab:parameters}). The red dashed curve corresponds to the diffuse gamma-ray flux without any absorption. The purple dot-dashed curve shows the diffuse flux after losses ($\gamma\gamma$ and Bethe-Heitler) and EM cascade at source. The final gamma-ray flux at Earth after EBL absorption  is represented by the thick red solid curve.  EBL absorption is more pronounced above $100$~GeV. Moreover, the propagation loss in the diffuse flux has attenuated the higher energy tail above $10^4$~GeV. Thus, the final diffuse flux peaks at around $10$~GeV and ends abruptly around $100$~GeV.  }
\label{with abs}
\end{figure}

Fig.~\ref{with abs} shows the   diffuse gamma-ray background for  Type IIn YSNe ($\zeta \approx 0.9$, see Fig.~\ref{fig:PiChart}) as a function of the observed photon energy.  
 The red-dashed curve shows the gamma-rays produced through $pp$ interactions (without any  energy loss). The purple dot-dashed curve shows the diffuse flux with source absorption and EM cascade, where  the small part of the flux at higher energies (above $10^3$ GeV) for the purple dot-dashed curve is due to the $\gamma\gamma$ cross-section as discussed in Sec.~\ref{subsec:source abs}. The red solid curve represents the gamma-ray flux after taking into account all the absorption (source+EBL) processes and EM cascades. The purple dot-dashed and red solid curves are  larger than the dashed red curve (no absorption) below $100$~GeV  because of the  additional  cascaded flux.

\begin{figure}[]
    \centering
      \includegraphics[width=0.49\textwidth]{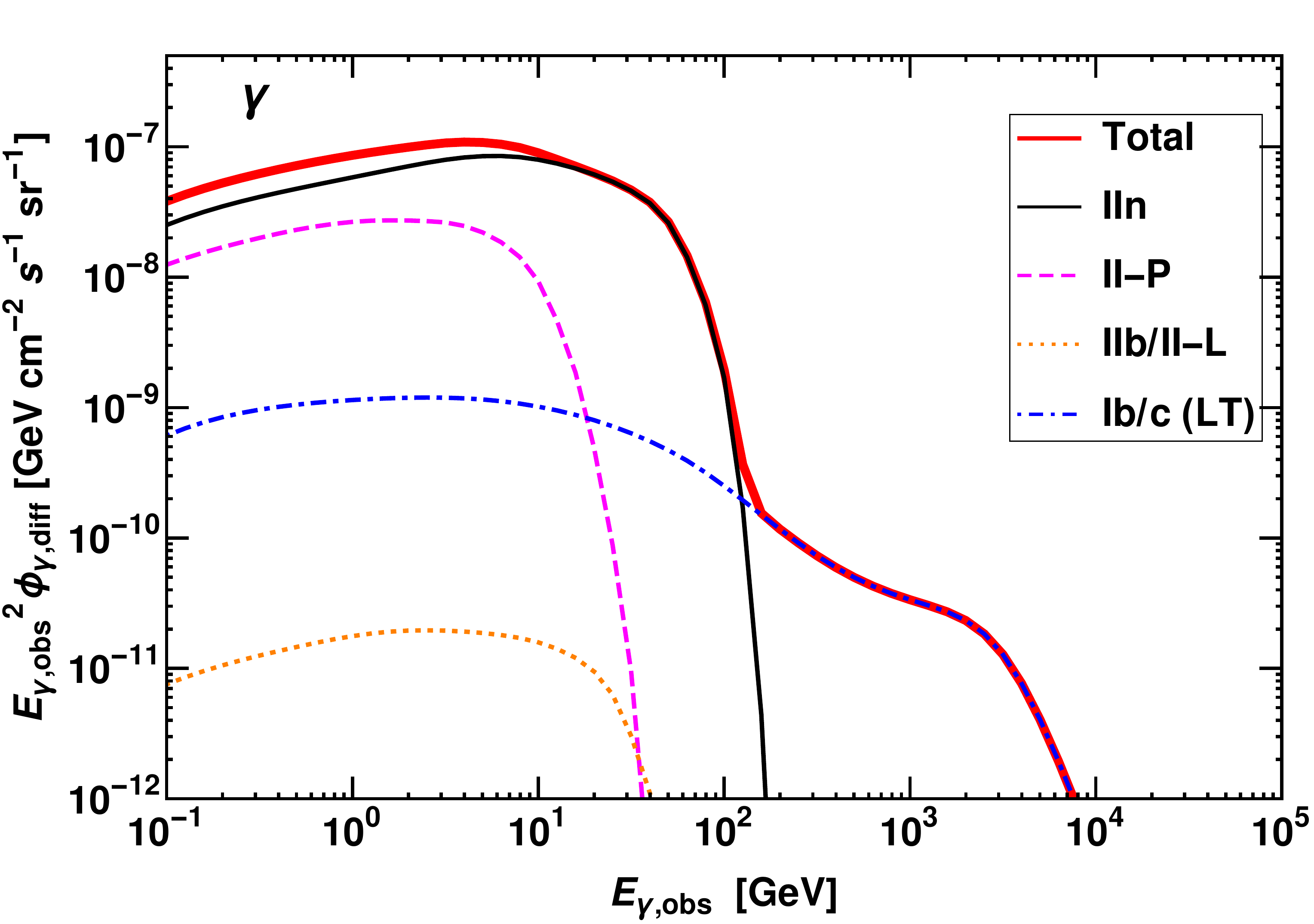}
      \includegraphics[width=0.49\textwidth]{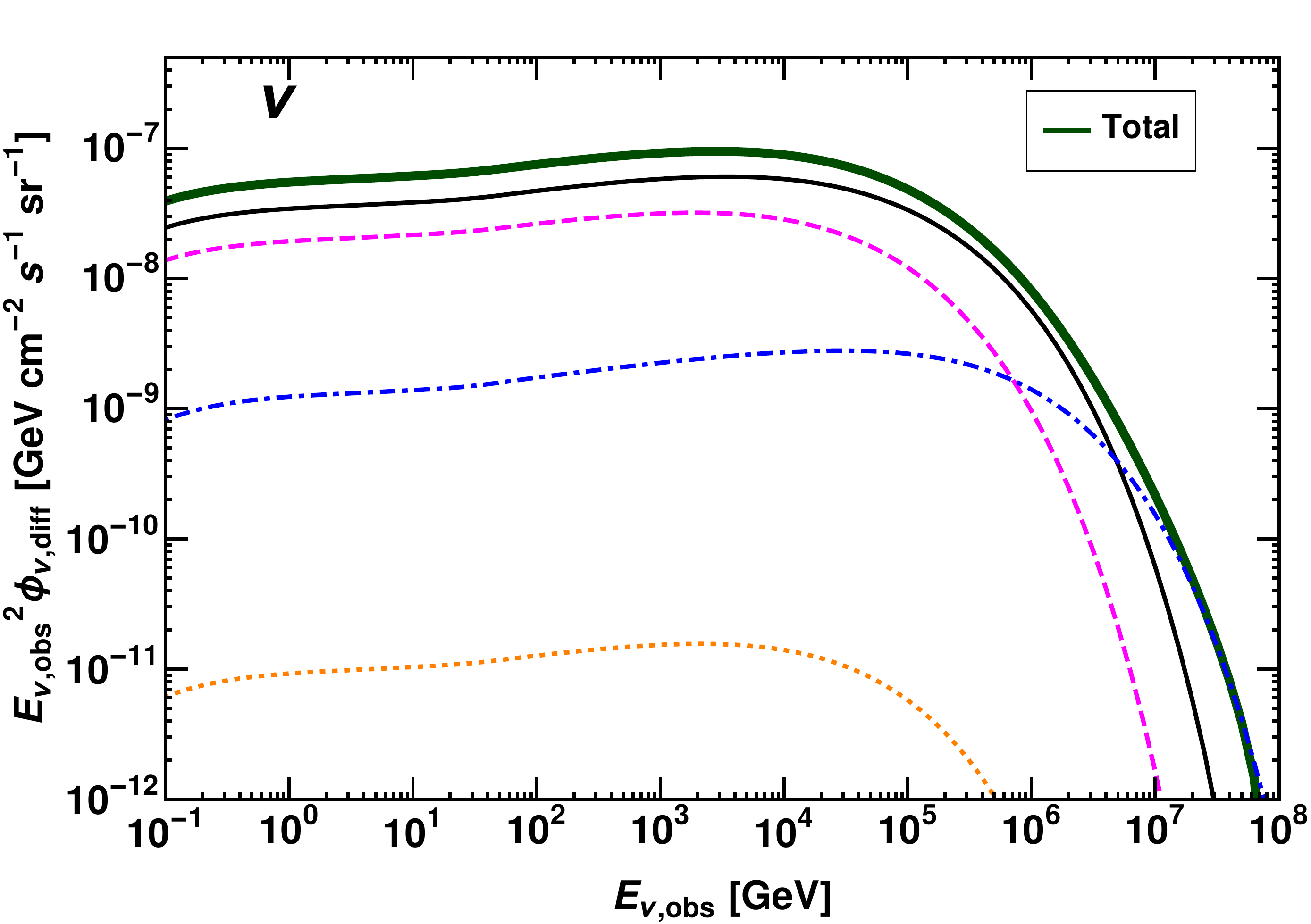}
    \caption{Diffuse  gamma-rays (left) and all-flavour neutrinos (right) fluxes as functions of the observed particle energy for the  different YSNe classes (see  Table \ref{tab:parameters} for our benchmark model parameters).
    For both  gamma-rays and neutrinos, Type IIn (black solid curves) YSNe contribute the most to the total  diffuse emission, followed by  Type II-P (magenta dashed curves)  YSNe  at lower energies. The contribution (despite being small) from Type Ib/c (LT) (blue dot-dashed curves) YSNe may show up above $10^2$~GeV in gamma-rays, whereas in neutrinos above $10^7$ GeV.   Type IIb/II-L (orange dotted curves) YSNe contribute negligibly. The flux of Type Ib/c YSNe is not shown  since it lies outside the plot range.   The total diffuse gamma-ray and neutrino fluxes are plotted as  thick red and thick green curves. It can be seen that IIn, II-P and Ib/c (LT) YSNe are the main contributors to the total diffuse spectra of both gamma-ray and neutrinos. }
    \label{fig:diff_et}
\end{figure}

The left panel of Fig.~\ref{fig:diff_et} displays the diffuse gamma-ray background for the different YSN Types by considering the benchmark values introduced in Table \ref{tab:parameters} and for  $\alpha_{\rm p}=2.0$.
The largest contribution to the diffuse emission comes from Type IIn YSNe. 
The diffuse flux of Type II-P YSNe falls at around $20$~GeV, i.e.~at lower energies than Type IIn YSNe because the thermal photons of Type II-P YSNe have larger average energies than the ones of Type IIn YSNe ($\epsilon_{av}=1$~eV for  IIn YSNe vs.~$\epsilon_{av}=4.6$ eV for II-P YSNe). The diffuse flux of Type Ib/c (LT) YSNe  falls at larger observed gamma-ray energies than the one of Type IIn YSNe because of the average energy of thermal photons. It has been found that Type II-P contributes small amount to the diffuse background at lower energies (below $10$~GeV).  Overall,  the contribution of Type IIb/II-L, and Ib/c (LT) YSNe to the total diffuse gamma-ray background is negligible. However, a small diffuse flux of gamma-rays from Ib/c (LT) YSNe may show up above $10^2$~GeV.  Therefore, the total gamma-ray background (thick red curve) is mostly dominated by the contribution from Type IIn SNe.

\subsection{Diffuse background of high-energy neutrinos}
The diffuse background of high-energy neutrinos for the different YSN Types can be computed by relying on Eq.~\ref{eq:diff}. The crucial difference with respect to the diffuse gamma-ray flux is the loss of the gamma-ray flux due to the propagation. 

The diffuse background of high-energy neutrinos is shown in the right panel of Fig.~\ref{fig:diff_et} for our benchmark model parameters introduced in Table~\ref{tab:parameters} and for  $\alpha_{\rm p}=2.0$.
Similar to what observed for gamma-rays (see left panel of Fig.~\ref{fig:diff_et}), the contribution  from Type IIn YSNe is  larger than the one  of Type II-P and Ib/c (LT) YSNe  below $E_{\nu}=10^7$~GeV.  However, Type II-P YSNe also have significant contribution and their flux is smaller than IIn YSNe by about a factor of $2$. The diffuse flux of Type Ib/c (LT) is small at lower energies but might show up above  $E_{\nu}=10^7$ GeV, this is due to the fact that the maximum proton energy of this YSN Type is large because of the larger shock velocity and shock radius (see also Fig.~\ref{fig:point_source}). Also note that the contribution of Type IIb/II-L YSNe to the total diffuse emission of high-energy neutrinos is  negligible as the source flux of Type IIb/II-L YSNe is already  quite small in comparison to the one of other Types of YSNe.

\subsection{Model parameter uncertainties}

\begin{figure}[]
    \centering
    \includegraphics[scale=0.4]{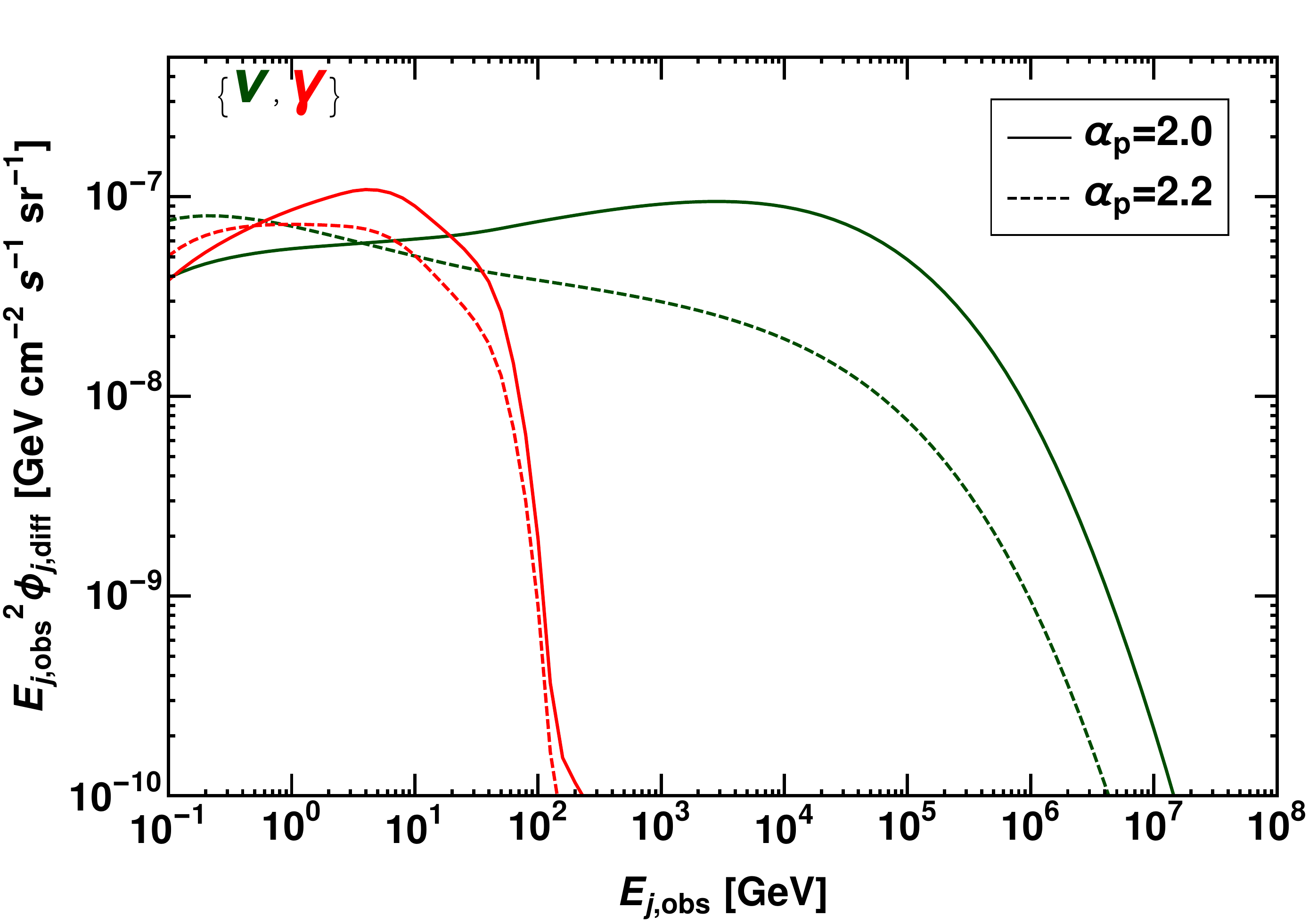}
    \caption{Diffuse backgrounds of gamma-rays (in red) and all-flavour neutrinos (dark green) for all YSNe Types as a function of the observed particle energy for  $\alpha_{\rm p}=2.0$ (solid) and $\alpha_{\rm p}=2.2$ (dashed). 
    The subscript $j$ represents $\nu$ and $\gamma$. 
    Larger values of $\alpha_{\rm p}$ are responsible for softer energy spectra for both gamma-rays and neutrinos; the dependence of the diffuse spectrum on  $\alpha_{p}$ is more pronounced for neutrinos than for gamma-rays.}
    \label{fig:All_nu_gamma}
\end{figure}
The injection spectral index of protons crucially affects  the high-energy diffuse backgrounds.
Fig.~\ref{fig:All_nu_gamma} shows the total diffuse gamma-ray  and neutrino backgrounds for $\alpha_{\rm p}=2.0$ (solid curves) and $\alpha_{\rm p}=2.2$ (dashed curves). 
Both diffuse backgrounds are larger for $\alpha_{\rm p}=2.0$ than that for $\alpha_{\rm p}=2.2$. This is due to the hardness of the energy distribution for  $\alpha_{\rm p}=2.0$ at higher energies. Moreover, for gamma-rays,  the EM cascaded flux of $\alpha_{\rm p}=2.0$ is larger than that obtained by using $\alpha_{\rm p}=2.2$. The dependence on  $\alpha_{\rm p}$ is more pronounced for neutrinos than for gamma-rays. This is because the higher energy part (above $100$~GeV) of  gamma-rays is heavily attenuated.

\begin{figure}[]
    \centering
      \includegraphics[width=0.49\textwidth]{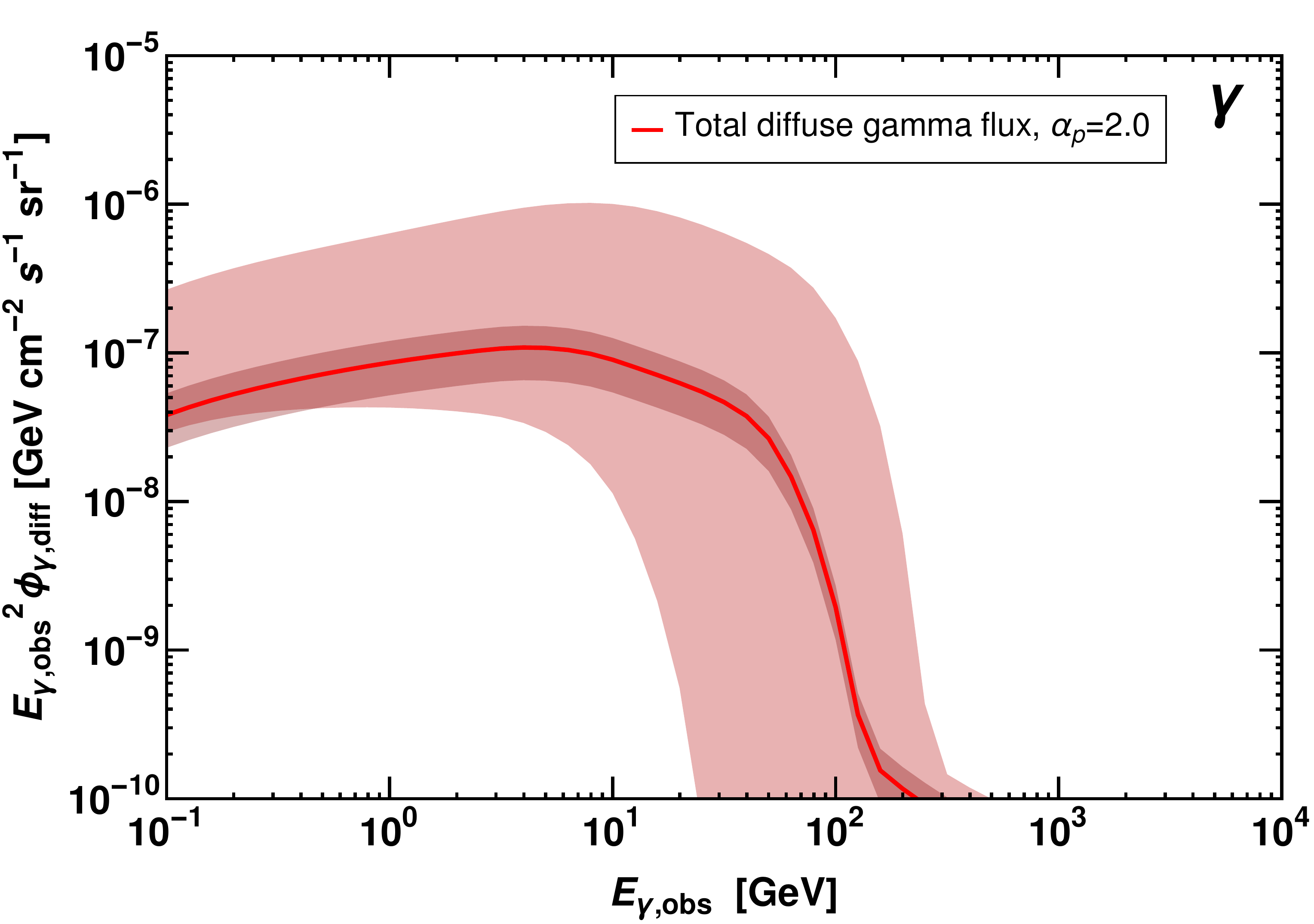}
      \includegraphics[width=0.49\textwidth]{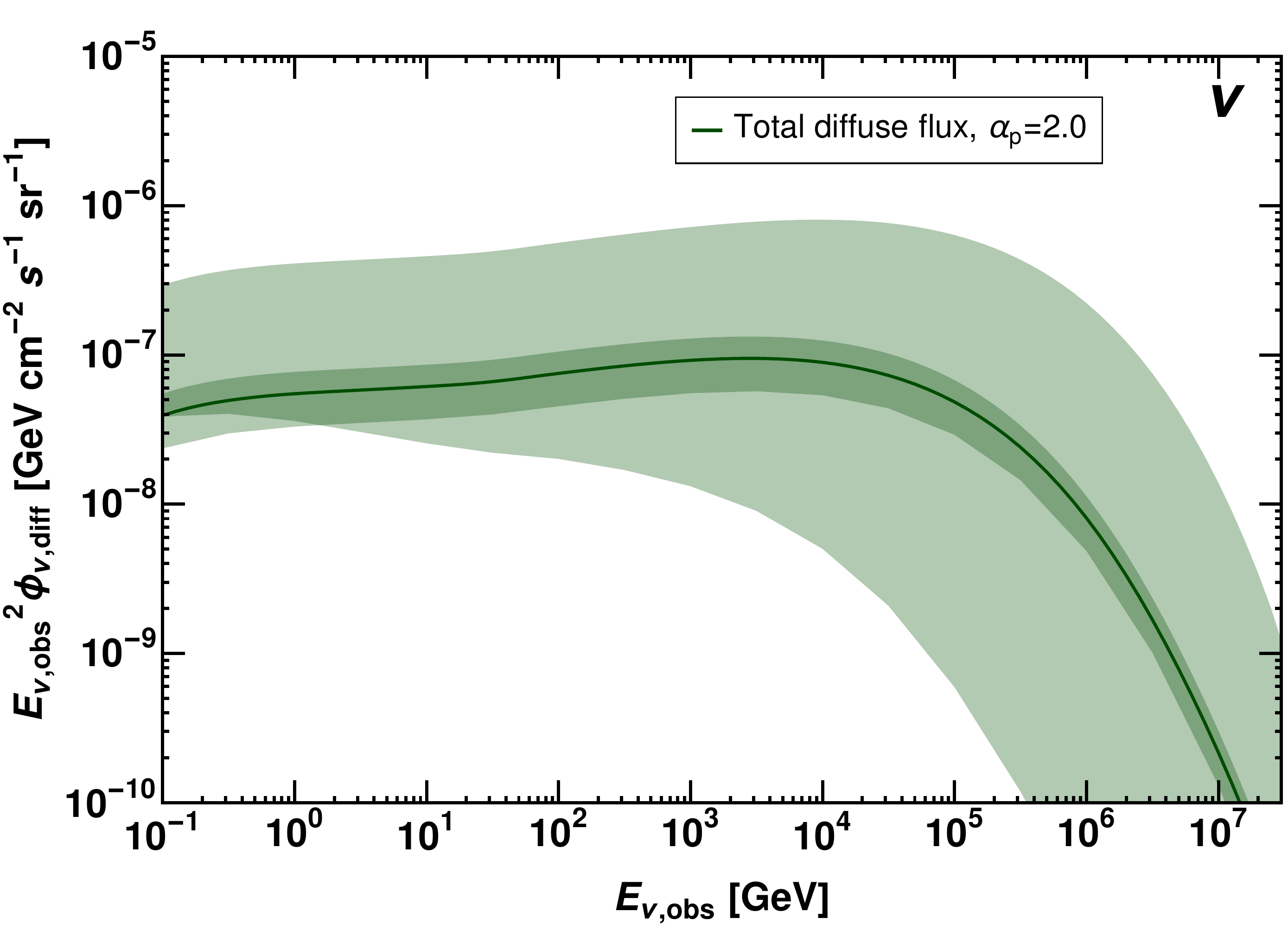}
    \caption{
    {\it Left panel:} Diffuse gamma-ray background from all Types of YSNe as a function of the observed gamma-ray energy. The solid red line indicated our benchmark diffuse emission for  $\alpha_{\rm p}=2.0$ and the other model parameters fixed as in Table~\ref{tab:parameters}. The band correspond to the uncertainties in shock velocity $v_{\rm sh}$, power law $\alpha_{\rm p}$, fraction of kinetic energy $\epsilon_{\rm p}$ and fraction of magnetic energy $\epsilon_{\rm B}$ (see Table~\ref{tab:uncertainty}). The uncertainty  associated to the local SN rate is represented by the thin band.
    {\it Right panel:}   Correspondent diffuse all-flavour neutrino background. 
     The uncertainty from model parameters is larger than  the one  from the SN rate.
    }
    \label{fig:tot_diff_nu}
\end{figure}

The left panel of Fig.~\ref{fig:tot_diff_nu} shows the total diffuse background of gamma-rays, including the contribution from all YSN Types.  For reference, the diffuse background estimated for benchmark model parameters as in Table~\ref{tab:parameters} and $\alpha_{\rm p}=2.0$ is represented by the solid red curve. The  uncertainty band results from the convolution of the uncertainties  on the parameters listed in Table \ref{tab:parameters}. 
In particular, the model parameters mostly affecting the spectral distribution are  $v_{\rm sh}$, $\epsilon_{\rm p}$ and $\epsilon_{\rm B}$; the upper and lower limits of the parameters are reported in Table~\ref{tab:uncertainty}. 
Our choices on the upper limits for these model parameters are  conservative compared to  
observations~\cite{Smith:2014txa,Yaron:2017umb,Bullivant:2018tru}. The remaining model parameters (see Table~\ref{tab:parameters}) are instead  kept fixed. Note that we have not considered uncertainties on the benchmark parameters of Type IIb/II-L YSNe because their contribution to diffuse  backgrounds is negligible. The lower limit of the gamma-ray diffuse emission is instead obtained for  $\alpha_{\rm p} = 2.2$ and  
 shows a softer energy distribution.  In addition, the uncertainty associated to the local SN rate (see Sec.~\ref{sec:diffuse}) might also have important consequences on the diffuse background spectra of gamma-rays and neutrinos. This uncertainty is included in our benchmark diffuse spectra and shown by the thin band. Interestingly, the uncertainty in the diffuse background from the SN rate is smaller than the one  from the model parameters. Analogously, the right  panel of Fig.~\ref{fig:tot_diff_nu} shows the total diffuse background of high-energy neutrinos.
\begin{table}[]
\caption{Uncertainties on a selection of model parameters of different YSN Types  (see Table~\ref{tab:parameters} for the benchmark parameters). Uncertainties on the model parameters of Type IIb/II-L and Ib/c YSNe are not included because their contributions to diffuse backgrounds of high-energy particles is negligible. 
}
\centering
\begin{tabular}{|c||c|c|c|c|c|c|}
\cline{1-7}
 {\bf Parameters} &  \multicolumn{2}{c|}{\bf Type IIn}  &  \multicolumn{2}{c|}{\bf Type II-P}   & \multicolumn{2}{c|}{\bf Type Ib/c (LT)}\\
\cline{2-7}& Upper & Lower  & Upper & Lower  & Upper  & Lower  \\
\hline
\hline
$v_{\rm sh}$ ($\rm kms^{-1}$) & $9.5 \times 10^3$ & $5 \times 10^3$ & $2\times10^4$ & $8\times 10^3$ &   $2\times10^4$ & $5\times 10^3$\\
\hline
$\epsilon_{\rm p}$ & 0.1 & 0.01 & 0.1 & 0.01 & 0.1 & 0.01 \\
\hline
$\epsilon_{\rm B}$ & $10^{-2}$ & $3 \times10^{-4}$ & $3\times10^{-2}$& $10^{-3}$ & $3\times10^{-2}$ & $10^{-2}$ \\
\hline
\end{tabular}
\label{tab:uncertainty}
\end{table}



\subsection{Discussion} \label{sec:discussion}
\begin{figure}[]
    \centering
    \includegraphics[scale=0.4]{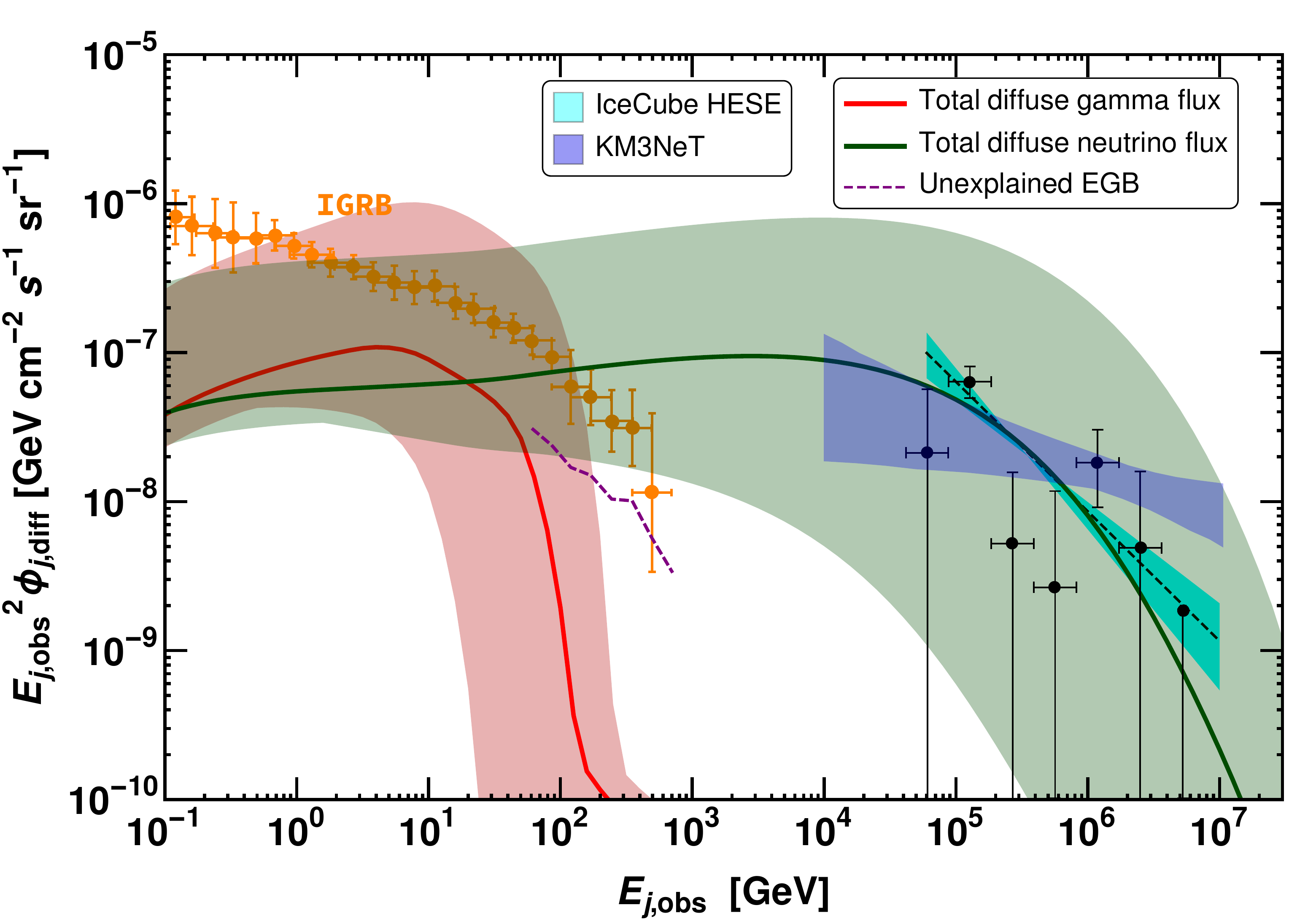}
    \caption{Total  diffuse gamma-ray (in red) and all-flavour neutrino (in green) backgrounds from YSNe as functions of the observed particle energy, analogous to Fig.~\ref{fig:tot_diff_nu}. 
    The subscript $j$ stands for $\nu$ or $\gamma$. For gamma-rays, the orange data points with error bars illustrate the diffuse gamma-ray background measured by Fermi-LAT (IGRB)~\cite{Ackermann:2014usa}. The purple dashed curve shows the unexplained portion of the IGRB~\cite{Fermi-LAT:2015otn,Fermi-LAT:2014ryh,Chakraborty:2015sta}. For neutrinos, the black-dashed line shows the IceCube (HESE) diffuse flux best fit for $7.5$~years of data (black data points with error bars); the cyan band depicts the  uncertainty on the IceCube diffuse flux at $68\%$ confidence level~\cite{IceCube:2020wum}. The diffuse flux sensitivity of the future neutrino experiment KM3NeT  is also shown by the light blue band \cite{KM3NeT:2021szv}. 
    It is evident that part of the parameter space considered for  YSNe is ruled out from  multi-messenger constraints from Fermi-LAT and IceCube. Nevertheless, our benchmark YSN  parameters (Table \ref{tab:parameters}) can very well explain part of  the IceCube diffuse flux  without the correspondent gamma-ray emission being in tension with  the Fermi-LAT gamma-ray data.   
    KM3NeT will further probe the  diffuse neutrino flux from YSNe in the energy range $10^{4}$-$10^{6}$~GeV.  
    }
    \label{fig:constrain}
\end{figure}

Fig.~\ref{fig:constrain} summarizes our findings on the gamma-ray and neutrino diffuse emission from YSNe. For reference, the data points of the  Fermi-LAT Isotropic Gamma-ray Background (IGRB) between $100$~MeV to $820$~GeV
are  shown~\cite{Ackermann:2014usa}. Moreover, the  unexplained IGRB component in the range $50$--$1000$~GeV is plotted with  the purple dashed line~\cite{Fermi-LAT:2014ryh,Chakraborty:2016mvc}. 
We also show the  data points with error bars corresponding to  $7.5$~years by IceCube  High-Energy Starting Event (HESE). 
The best fit to the IceCube data is represented by 
 the black-dashed line; the related $68\%$ confidence level uncertainty  is plotted in cyan~\cite{IceCube:2020wum}. The diffuse neutrino flux sensitivity at $68\%$ confidence level for  KM3NeT  is  shown in light blue~\cite{KM3NeT:2021szv}.

As for gamma-rays, a large fraction of the IGRB observed by Fermi-LAT may be coming  from  blazars, although 
 recent work shows evidence for   star-forming galaxies  as the dominant contributors to the  IGRB~\cite{Stecker:2010di,DiMauro:2017ing,Lisanti:2016jub,Linden:2016fdd,Roth:2021lvk}.
  Our findings are in agreement with this picture on the IGRB composition. 
In fact, our benchmark YSN gamma-ray background (red solid line)  is severely attenuated above $100$~GeV and  not in  tension with blazar unexplained flux (purple dashed line). 
Moreover, the  gamma-ray diffuse emission from star-forming galaxies should originate from the collisions of the SN accelerated protons with molecular clouds (ISM) in these active galaxies~\cite{Senno:2015tra,1996SSRv...75..279V} and
therefore  include the  contribution of YSNe as well. However, the gamma-rays created in YSNe undergo larger attenuation (due to the dense CSM environment) than gamma-rays created in a thin ISM \cite{Roth:2021lvk}. 
By comparing the diffuse gamma-ray emission predicted in this work with the Fermi-LAT data in  Fig. \ref{fig:tot_diff_nu}, it is evident that our benchmark diffuse gamma flux is smaller than the Fermi-LAT IGRB  and thus might  negligibly contribute  to the total SBG flux.

As for high-energy neutrinos, 
 our benchmark YSN neutrino background is in good agreement with the IceCube HESE data below $10^6$~GeV. 
Intriguingly, the YSN neutrino background at low energies ($10^4$--$10^5$~GeV) is significantly large and can explain the HESE   concentration at these energies, in alternative to dark sources or hypernovae~\cite{Murase:2015xka,Denton:2018tdj,Ahlers:2014ioa,Murase:2019vdl,Senno:2015tra}. Within the YSN interpration, the low-energy component of the neutrino diffuse background may originate from  Type IIn SNe, whereas  the neutrino diffuse background
 above $10^5$~GeV may also have contributions from II-P and Ib/c (LT) SNe. This multi component interpretation, in addition to likely contributions to the neutrino diffuse emission  from other sources,   can accommodate possible different power laws in different energy ranges  in future data fits
 of IceCube data~\cite{IceCube:2020wum}. The KM3NeT sensitivity shows that it will be able to probe the diffuse flux of neutrinos from YSNe in the energy range  $10^4$--$10^6$~GeV.

The contribution of YSNe may be subleading (e.g.~for our benchmark model parameters) to the IGRB, however it could explain very well the observed IceCube diffuse emission below $10^6$~GeV, relaxing the tension between gamma-ray and neutrino data for hadronic sources invoked as motivation for ``hidden'' sources~\cite{Murase:2015xka}.
Interestingly, part of the YSN parameter space allowed by multi-wavelength electromagnetic observations (see  Table~\ref{tab:uncertainty}) may overshoot the Fermi-LAT and IceCube data, as shown by the  bands in Fig.~\ref{fig:constrain}. This suggests that YSNe with such extreme model parameters are not representative of the YSN population.

The green and red solid curves in Fig.~\ref{fig:constrain} have been obtained for  $\epsilon_{\rm p}=0.01$ for Type IIn SNe. The maximum value allowed by the IGRB data is $\epsilon_{\rm p} \simeq 0.03$, for which  the corresponding neutrino background is  enhanced by a factor of $3$. Kinetic
simulations predict  $\epsilon_{\rm p}$ as large as $0.2$~\cite{Caprioli_2014}, which may be in conflict with the high-energy diffuse backgrounds, if representative of the whole Type IIn YSN population.

The diffuse neutrino flux from Type IIn SNe reported in Ref.~\cite{Petropoulou:2017ymv} relies on $\epsilon_{\rm p} \sim0.2$, which 
may be in tension with the Fermi-LAT IGRB. 
On the other hand, the Monte-Carlo simulations of Ref.~\cite{Petropoulou:2017ymv}  for randomly distributed $\epsilon_{\rm p}$ in the range $[0.01, 0.1]$ produced a smaller diffuse flux, contributing up to  $10\%$ to  the IceCube HESE flux.
We find that the IceCube HESE data can be accommodated by smaller  values of $\epsilon_{\rm p}$, thus remaining consistent with the IGRB data. In our case, the smaller $\epsilon_{\rm p}$ is
allowed as the total diffuse background  also includes non-negligible contributions from Type IIP and Ib/c (LT) YSNe. In particular,  Ib/c (LT) YSNe 
having the harder spectra  dominate the higher energy tail above $1$~PeV.

Our findings are also in agreement with the ones of Ref.~\cite{Fang:2020bkm}, which estimated the neutrino and gamma-ray emission from a range of non-relativistic shock powered transients, concluding that the observed neutrino emission may come from gamma-ray dim sources. 
However,  their model is based on  optical observations of these transients, while we relied on YSN model parameters coming from a 
wide range of multi-wavelength surveys.

Upper limits on the  high-energy diffuse neutrino  background from SNe have been provided in 
Refs.~\cite{phdthesis,IceCube:2021oiv}, which   found that the diffuse neutrino background is dominated by SNe II-P,  followed by SNe IIn and Ib/c. These findings are in contrast with ours because of the  different choices of the local SN rates (Ref.~\cite{phdthesis} assumes $\xi= 52.4\%$ for SNe II-P, $6.4\%$ for SNe IIn, and $25\%$ for SNe Ib/c, while we considered the YSN fractions summarized in Fig.~\ref{fig:PiChart}). In addition, Refs.~\cite{phdthesis,IceCube:2021oiv} do not take into account  gamma-ray constraints.

We note that 
 the larger mass loss rates of SNe II-P reported in Refs.~\cite{Morozova:2016asf,Morozova:2016efp} (see also  
 Sec.~\ref{sec:DOSNR}) can produce a diffuse emission from YSN II-P   comparable to that of Type IIn SNe. This, in turn, may further constrain  the parameters characteristics of the SN IIn population.

\section{Detection prospects of nearby young supernovae in  gamma-rays and neutrinos}
\label{sec:Point_source_detection}

\begin{figure}[]
    \centering
  \vbox{
  \hbox{
  \includegraphics[width=0.48\textwidth]{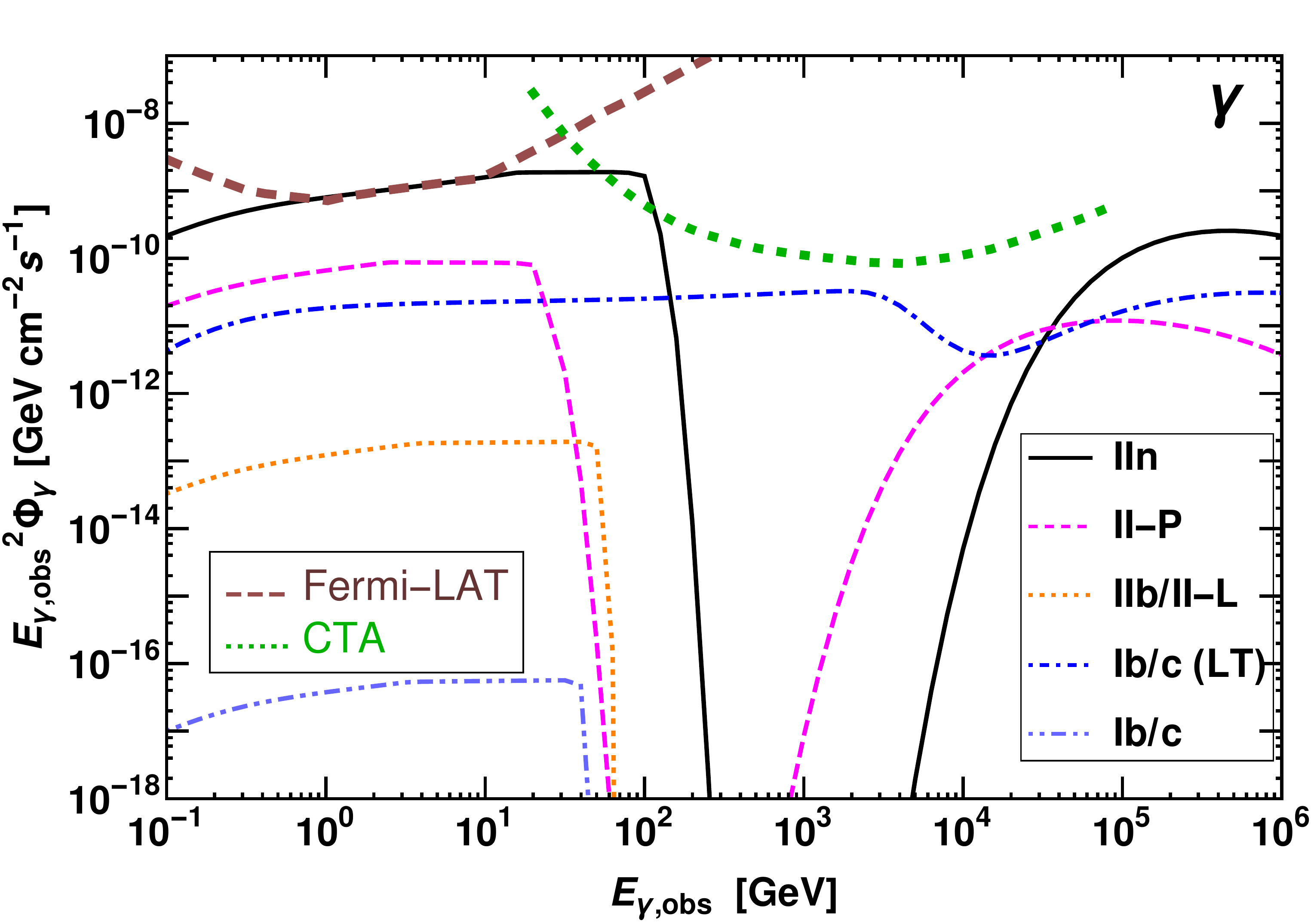}
  \includegraphics[width=0.48\textwidth]{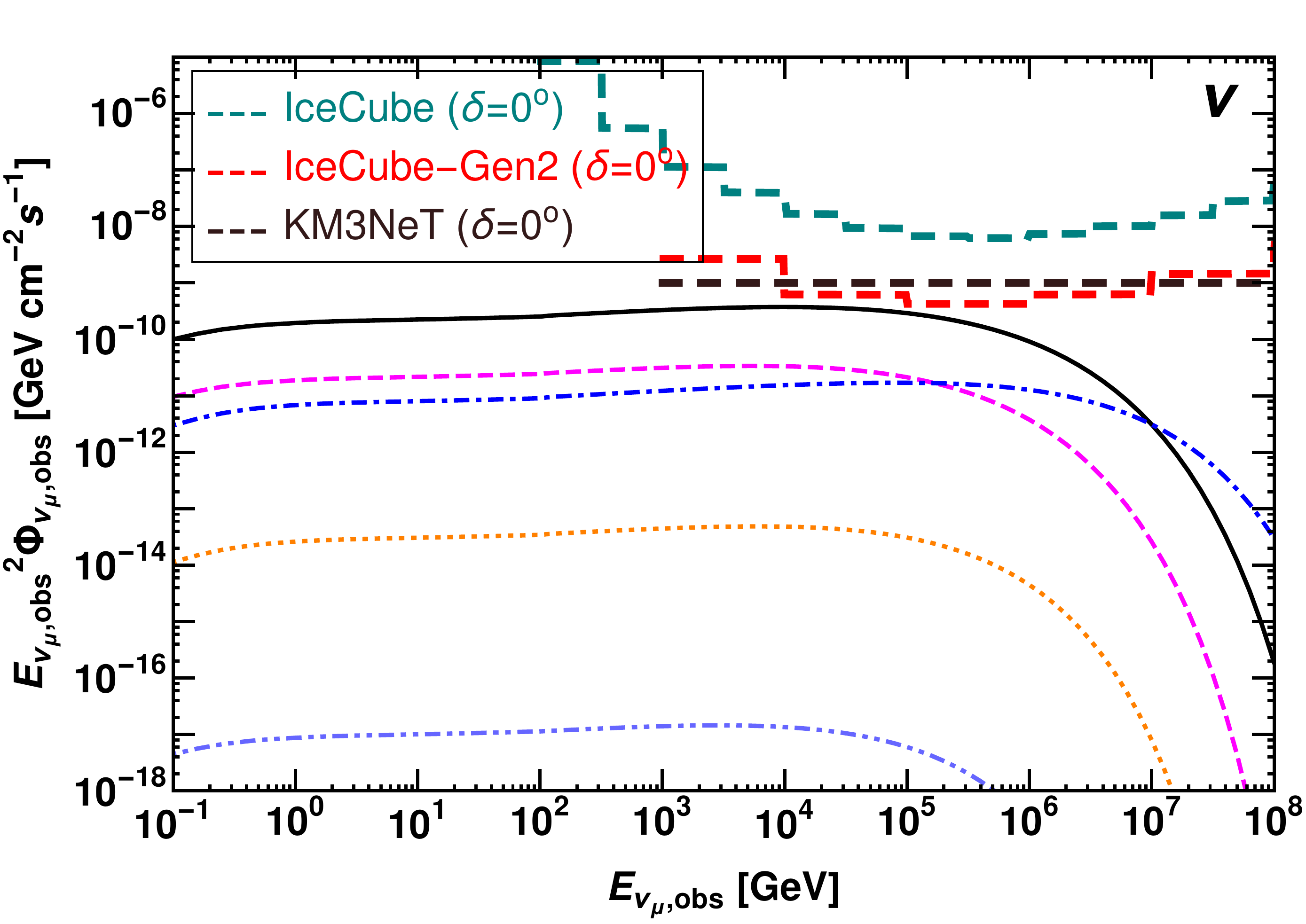}
  }
  \vspace{0.2cm}
  \hbox{
    \includegraphics[width=0.48\textwidth]{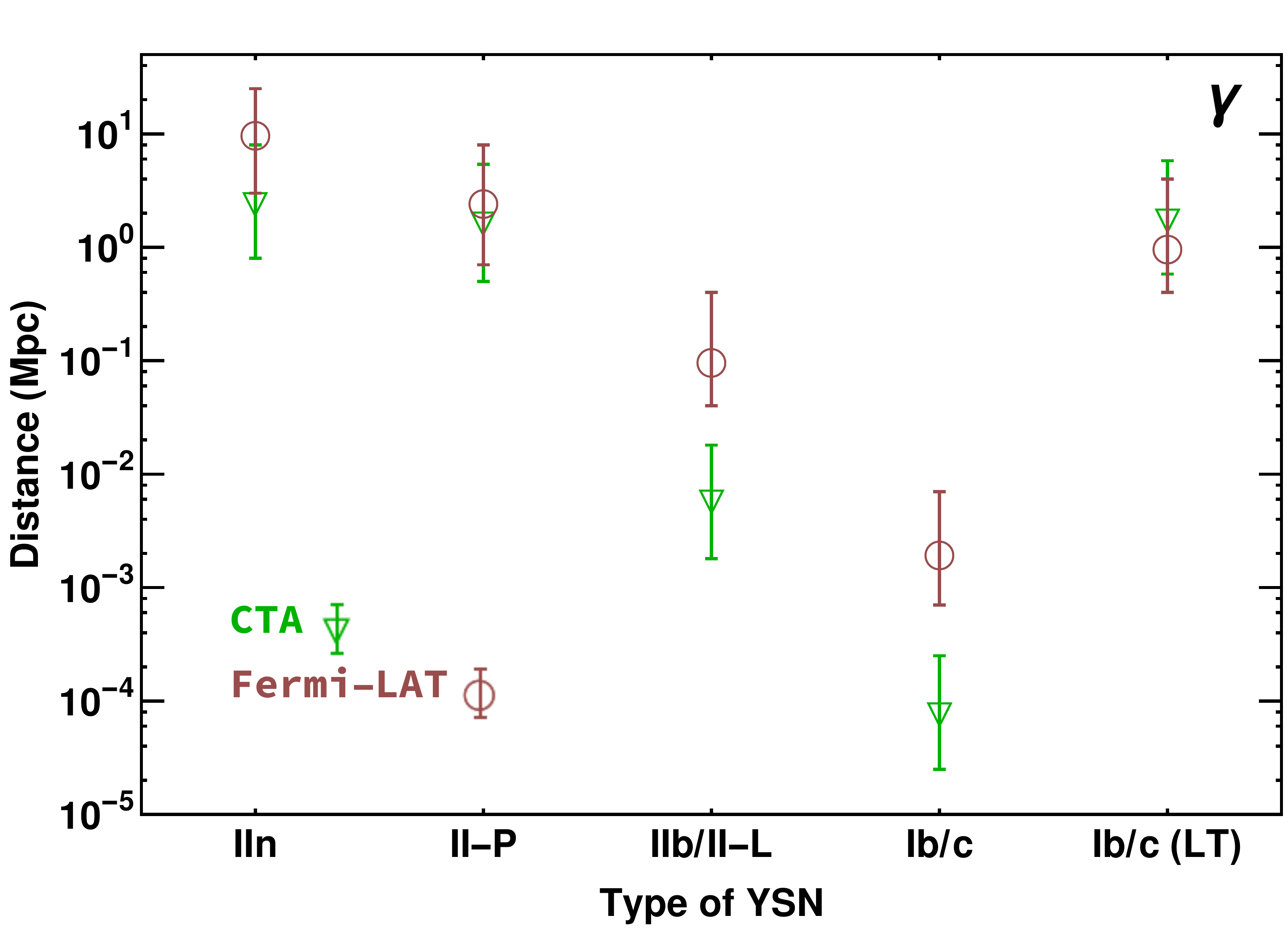}
  \includegraphics[width=0.48\textwidth]{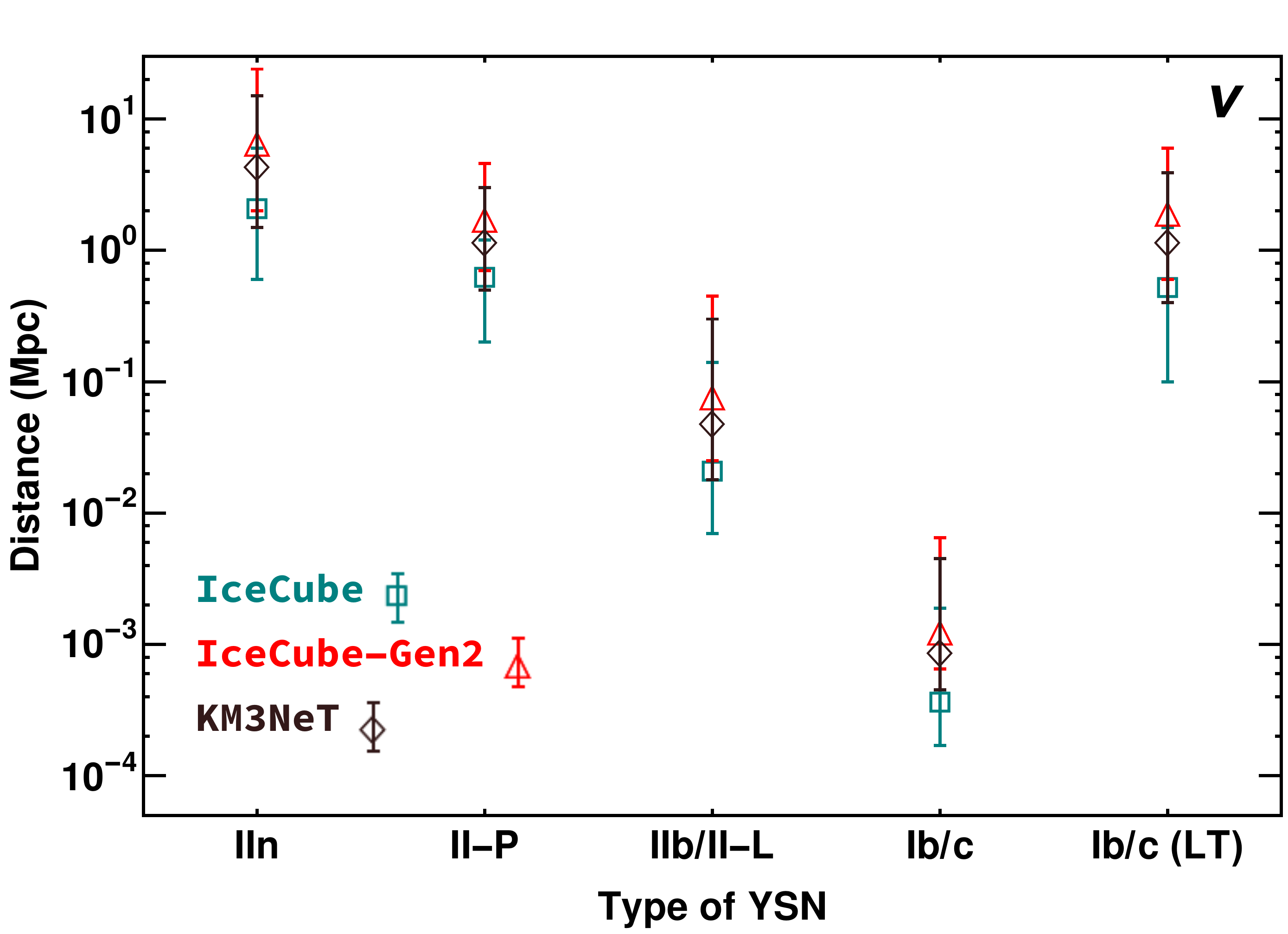}
  }
  }
    \caption{Detection prospects of nearby supernovae  using gamma-ray  (left) and neutrino
    telescopes (right). {\it Top left panel:} Gamma-ray energy fluxes from the different YSN Types  at $10$~Mpc  as functions of the observed particle energy. The one year Fermi-LAT sensitivity is shown by the thick light brown dashed curve \cite{Fermi-LAT:Sensitivity}
    and the thick green dotted curve represents the $100$ hour CTA sensitivity~\cite{2011ExA....32..193A}.  Type IIn YSNe may be detected by both Fermi-LAT and CTA, while all other sources will be too dim at 10 Mpc.  {\it Top right panel:} Corresponding muon neutrino  energy fluxes. 
    The sensitivities of IceCube, IceCube-Gen2 and KM3NeT  for point source detection  are plotted (thick dashed lines) in dark cyan (IceCube) \cite{IceCube:2016tpw}, red (IceCube-Gen2) \cite{IceCube-Gen2:2020qha} and dark brown (KM3NeT) \cite{KM3NeT:2018wnd}. The sensitivities of IceCube, IceCube-Gen2 and KM3NeT are plotted for the declination angle $\delta=0^{o}$. All these neutrino observatories will be able to detect YSNe at distances smaller than $10$ Mpc. 
    {\it Bottom left panel:} Gamma-ray YSN detection horizon for   Fermi-LAT (light brown) and CTA (green)    as  functions of the YSN Type. For each YSN Type, the error band takes into account the model uncertainties (see Sec.~\ref{sec:discussion}).  Fermi-LAT and CTA could detect YSNe up to  $10$~Mpc  (see YSNe IIn);   CTA could have better sensitivity than Fermi-LAT and reach up to $2$ Mpc for YSNe Ib/c (LT). {\it Bottom right panel:} Corresponding  neutrino YSN detection horizon for IceCube (dark cyan), IceCube-Gen2 (red) and KM3NeT (dark brown).  IceCube-Gen2 will be able to detect YSNe  up to $\sim 4$ Mpc (see YSNe Type IIn). 
    }
    \label{fig:DetectionPoint}
\end{figure}

As shown in the previous Section, the diffuse backgrounds of neutrinos and gamma-rays from YSNe 
have  large uncertainties due to the widely varying model parameters. The detection of neutrinos and gamma-rays from nearby YSNe will help to further constrain these model parameters and can potentially provide complementary  understanding of shock-CSM interactions.  In the following, we investigate the detection prospects of  YSNe with present and future gamma-ray telescopes and neutrino detectors.

Figure~\ref{fig:DetectionPoint} illustrates the detection prospects for our benchmark YSNe in gamma-rays and neutrinos. 
The top panels display the energy flux expected at Earth for a YSN at $10$~Mpc. The top left panel shows that the gamma-ray energy flux would be detectable for Type IIn YSN by  Fermi-LAT~\cite{Fermi-LAT:Sensitivity} and CTA \cite{2011ExA....32..193A}, while other SN Types can be probed at distances smaller than $10$~Mpc. Interestingly, due to gamma-ray attenuation, dips could appear in the gamma-ray spectra between $100$--$1000$ GeV (see  related discussion for Fig.~\ref{source gamma}), CTA may be able to probe this feature for local YSNe. However, the detection of such dips requires more detailed analysis~\cite{Cristofari:2022low,Cristofari:2020pzg}.
The top right panel shows that corresponding neutrino flux from all YSNe  are detectable in IceCube~\cite{IceCube:2016tpw} below $10$~Mpc, in agreement with  Ref.~\cite{Murase:2017pfe}.  Km3NeT \cite{KM3NeT:2018wnd} and IceCube-Gen2 \cite{IceCube-Gen2:2020qha} will also be able to detect neutrinos from these sources below $10$~Mpc.

The bottom panels of Fig.~\ref{fig:DetectionPoint} show the detection horizon of existing and upcoming gamma-ray and neutrino telescopes for the different YSN Types, by considering the model parameter uncertainties, see Table \ref{tab:uncertainty}.
The reach of different telescopes is calculated by taking into account the integrated flux of each YSN in the specific energy range of detectors' best integral sensitivity. For gamma-rays, the energy range [$10^{-1},10^1$] GeV is chosen for Fermi-LAT as it has the best integral sensitivity in this range. However, the whole energy range of CTA, i.e., [$2\times10^1,7\times10^4$]~GeV is taken into account as the fluxes of most of  YSNe are attenuated where CTA is  most sensitive. 
The maximum reach of Fermi-LAT and CTA  for different YSNe is shown in the bottom left panel of Fig.~\ref{fig:DetectionPoint}. Fermi-LAT and CTA could detect YSNe up to  $10$~Mpc  (see YSNe IIn, for the benchmark parameters). However, Fermi-LAT has comparatively better sensitivity to YSNe than CTA except for Ib/c (LT) YSNe. This is due to the fact that CTA is more sensitive at higher energies and  gamma-rays at these energies are attenuated  by the thermal photons in the source. The exception of Ib/c (LT) YSNe is accounted for the small gamma-ray attenuation in the source (see top left plot and the discussion of Fig.~\ref{fig:point_source}).  The detection horizon of CTA is affected by the amount of gamma-ray attenuation by the source thermal photons as discussed above. Thus, the detection of these sources in gamma-rays   has the potential to unleash information about particle production, emission and propagation in YSNe. 

Similarly, for the neutrino telescopes, we consider the YSNe fluxes in the energy range [$10^5,10^6$]~GeV as these detectors have best sensitivity to the astrophysical neutrino flux in these energies (see the top right panel). The detection horizons marked in the plot refer to the ones for which  the integrated flux is equal to the detectors' integral sensitivity (see the top  panels).  For example, for YSNe Type IIn at distances above $2$ Mpc, the integrated neutrino flux in the energy range $10^5$--$10^6$ GeV would  fall below the IceCube sensitivity at $\delta = 0^o$ declination. The error bars represent the uncertainties in the model parameters. 
Our findings concerning the detectability of neutrinos from YSNe  IIn  are in agreement with the ones of  Ref.~\cite{Petropoulou:2017ymv,Marcowith:2014kfa}.
Considering the uncertainties in the model parameters, IceCube-Gen2 will instead be able to detect Type IIn YSNe  up to $\sim 20$ Mpc ($4$ Mpc for the benchmark parameters).

\section{Conclusions}
\label{sec:conclusion}

In this paper, we investigate the   high energy neutrino and gamma ray signals from different classes of young SNe (Type IIn, II-P, IIb/II-L and Ib/c YSNe), up to one year after their explosion. During this time,   shock accelerated protons interact with the dense CSM around these objects, leading to the production of secondary high-energy neutrinos and gamma-rays. In particular, for the first time, we also investigate the  late time emission of YSNe  Ib/c, coming from  the interaction of the SN shock with a dense hydrogen rich CSM far away from the stellar envelope.

Despite intense research activity, the  origin of the bulk of the diffuse high-energy neutrino background observed by the IceCube Neutrino Observatory as well as the the Isotropic Gamma-Ray Background detected by Fermi-LAT is yet unknown. Our benchmark (intermediate) diffuse neutrino emission is in excellent agreement with the  IceCube HESE data and suggests that YSNe could constitute the bulk of the high-energy diffuse neutrino emission observed by IceCube below $10^6$~GeV, while being dim enough in gamma-rays to avoid any conflict with Fermi-LAT data.

The largest contribution to the diffuse neutrino emission mainly comes from  Type IIn YSNe, followed by II-P and Ib/c (LT) YSNe, with Type IIn YSNe dominating the overall neutrino diffuse emission up to $10^7$~GeV. Type Ib/c (LT) YSNe populate the diffuse neutrino emission above $10^7$ GeV. The Type II-P YSN contribution is also significant, but smaller than  the Type IIn SN one. Similar findings also hold for the diffuse gamma-ray  background, after including all attenuation effects taking place both in the source and en route to Earth.

Intriguingly, the uncertainty bands obtained for the diffuse high-energy backgrounds by taking into account the uncertainties on the shock velocity, injection spectral index of protons,  the kinetic and magnetic energy fractions, and the YSN rate suggest that a large fraction of the YSN parameter space inferred from multi-wavelength electromagnetic observations of YSNe is excluded by Fermi-LAT and IceCube HESE data.

The  detection prospects of nearby YSNe with existing and upcoming gamma-ray  and neutrino telescopes have also been explored. Among all YSNe,  Type IIn SNe have the best discovery potential up to $\sim 10$~Mpc ($\sim 4$~Mpc)   in gamma-rays with Fermi-LAT and CTA (in  neutrinos with IceCube, KM3NeT and IceCube-Gen2). Interestingly, CTA may be able to distinguish gamma-ray attenuation features in the spectral energy distributions for nearby transients. However, a possible amplification of magnetic field, e.g.~due to Bell non-resonant streaming of cosmic rays, might influence  proton acceleration and the gamma-ray spectra~\citep[see e.g.,][]{10.1111/j.1365-2966.2004.08097.x,Zirakashvili:2008lwi,Haggerty:2019anu}. Multimessenger observations of such point sources will be able to probe these mechanisms and requires future dedicated work.

To conclude,  the high-energy neutrino emission (especially coming from Type IIn, II-P and Ib/c (LT) SNe)  can be a strong contender to explain the low-energy  IceCube HESE dataset. 
The corresponding 
diffuse gamma ray emission  is not in tension with the  IGRB data from Fermi-LAT  as  
YSNe gamma-rays are heavily attenuated and effectively ``hidden.'' The detection of high-energy particles from young supernovae will provide new insights on the processes linked to particle acceleration in young SNe.

\acknowledgments
We are grateful for helpful discussions with Madhurima Chakraborty, Jens Hjorth, Raffaella Margutti, and especially  Tetyana Pitik. S.C acknowledges the support of the Max Planck India Mobility Grant from the Max Planck Society, supporting the visit and stay at MPP during the project. S.C has also received funding from DST/SERB projects  CRG/2021/002961 and MTR/2021/000540. This project has received funding from the  Villum Foundation (Project No.~37358), the Carlsberg Foundation (CF18-0183), the Deutsche Forschungsgemeinschaft through Sonderforschungbereich
SFB~1258 ``Neutrinos and Dark Matter in Astro- and Particle Physics'' (NDM). Parts of this research were supported by the Australian Research Council Centre of Excellence for All Sky Astrophysics in 3 Dimensions (ASTRO 3D), through project number CE170100013. 

\appendix
\label{appendix}
\section{Characteristic timescales for proton acceleration and radiative processes} \label{sec:appendix}

 Shock accelerated protons lose energy through different processes such as $pp$ interactions, dynamic or adiabatic losses, proton synchrotron, inverse Compton, Bethe-Heitler, etc. In this Appendix, we provide an overview of the different energy loss time scales for high energy protons and compare them with the acceleration time scales. 
 
 \begin{figure}[]
    \centering
    \vbox{
      \hbox{\includegraphics[width=0.47\textwidth]{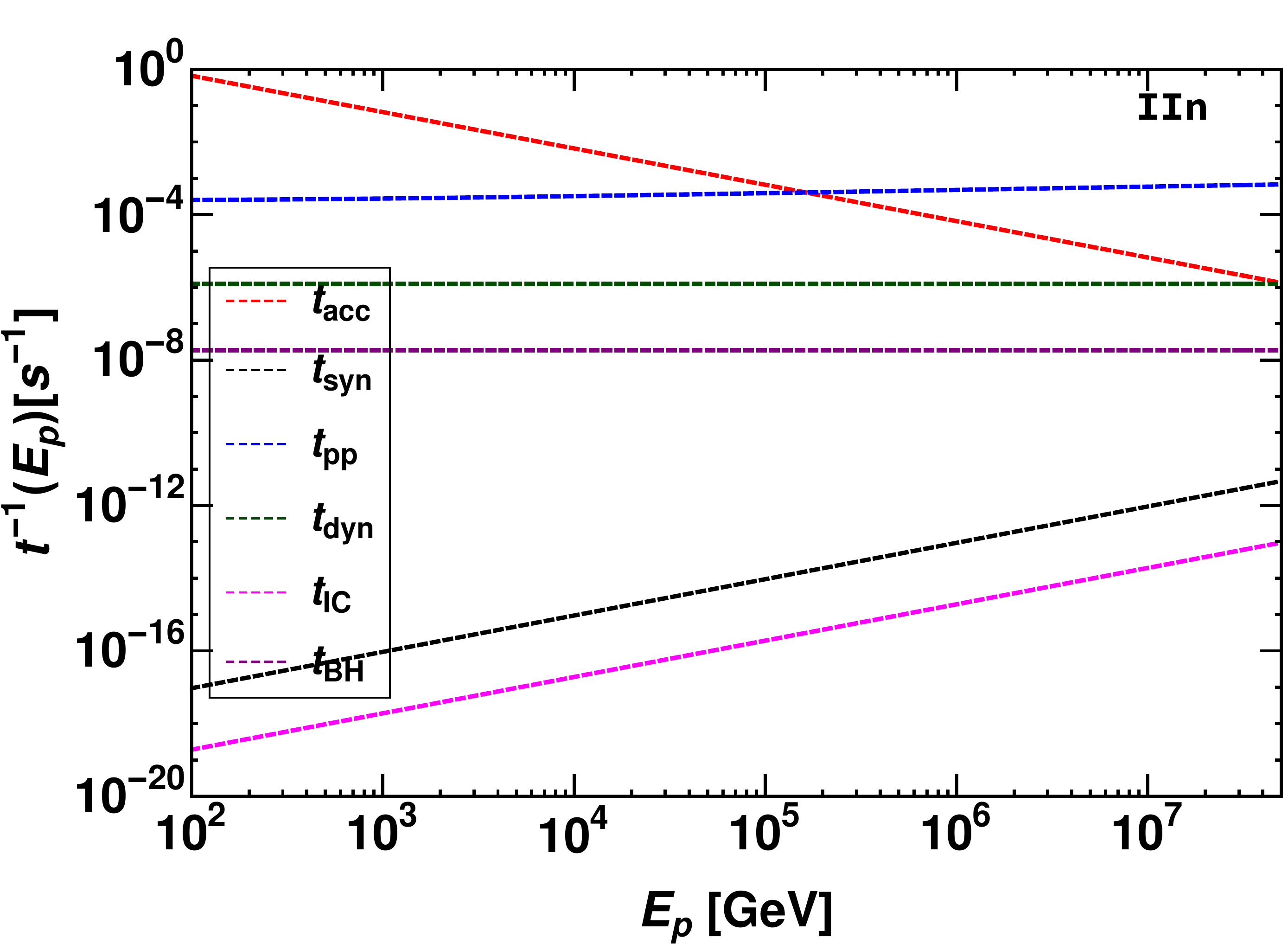}\hspace{0.5cm}\includegraphics[width=0.47\textwidth]{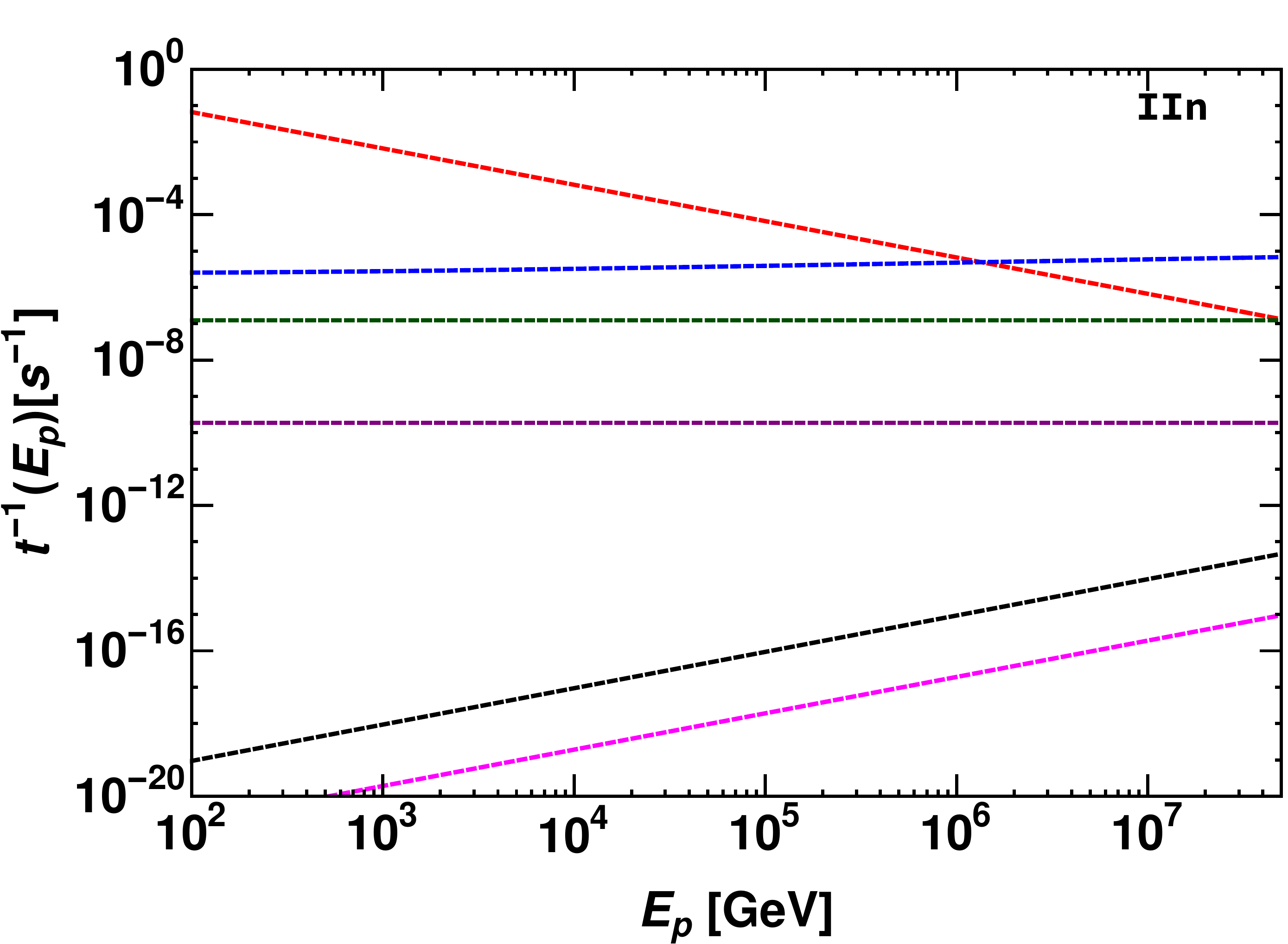}}
      \vspace{0.5cm}
      \hbox{\includegraphics[width=0.47\textwidth]{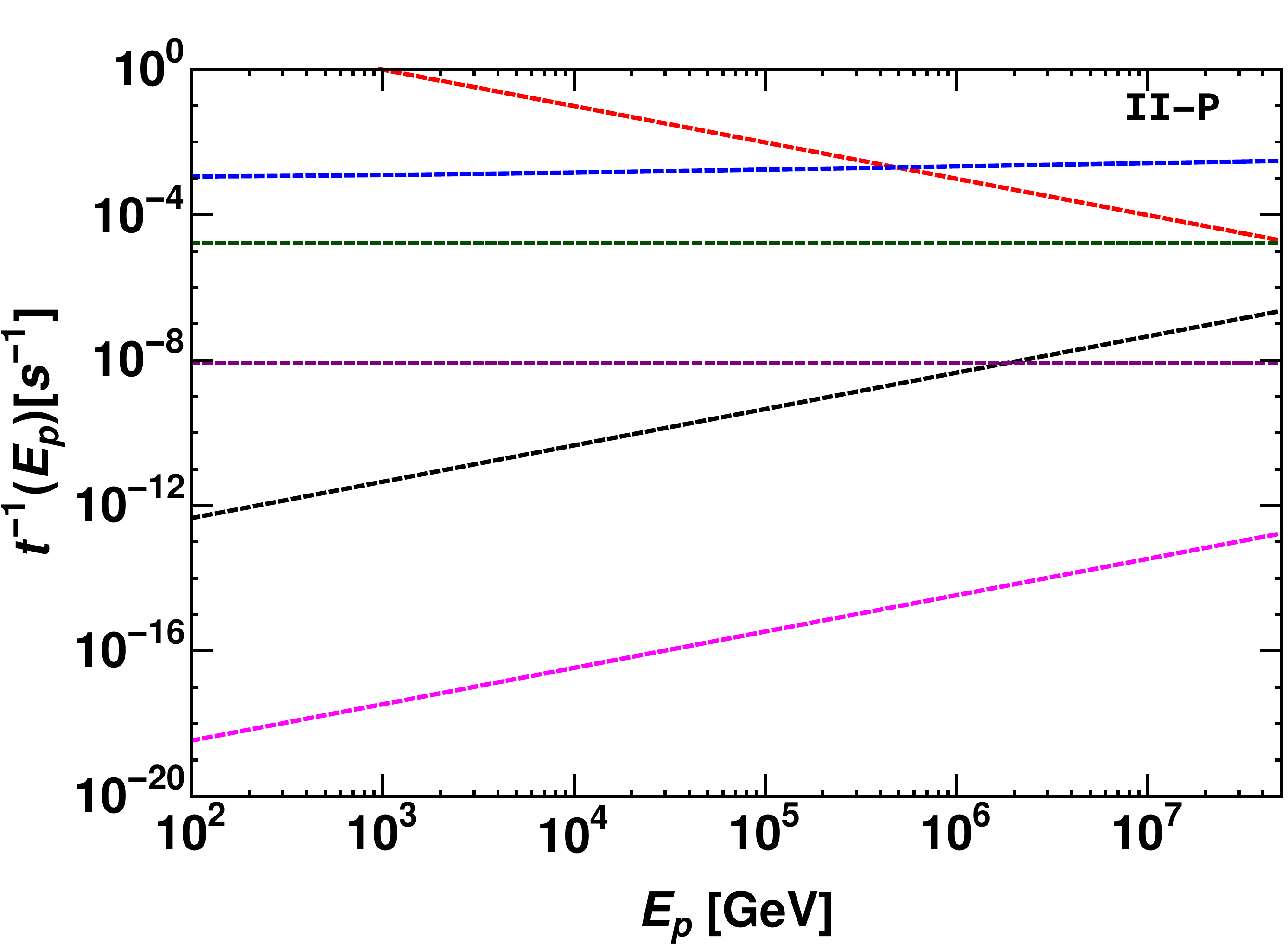}\hspace{0.5cm}\includegraphics[width=0.47\textwidth]{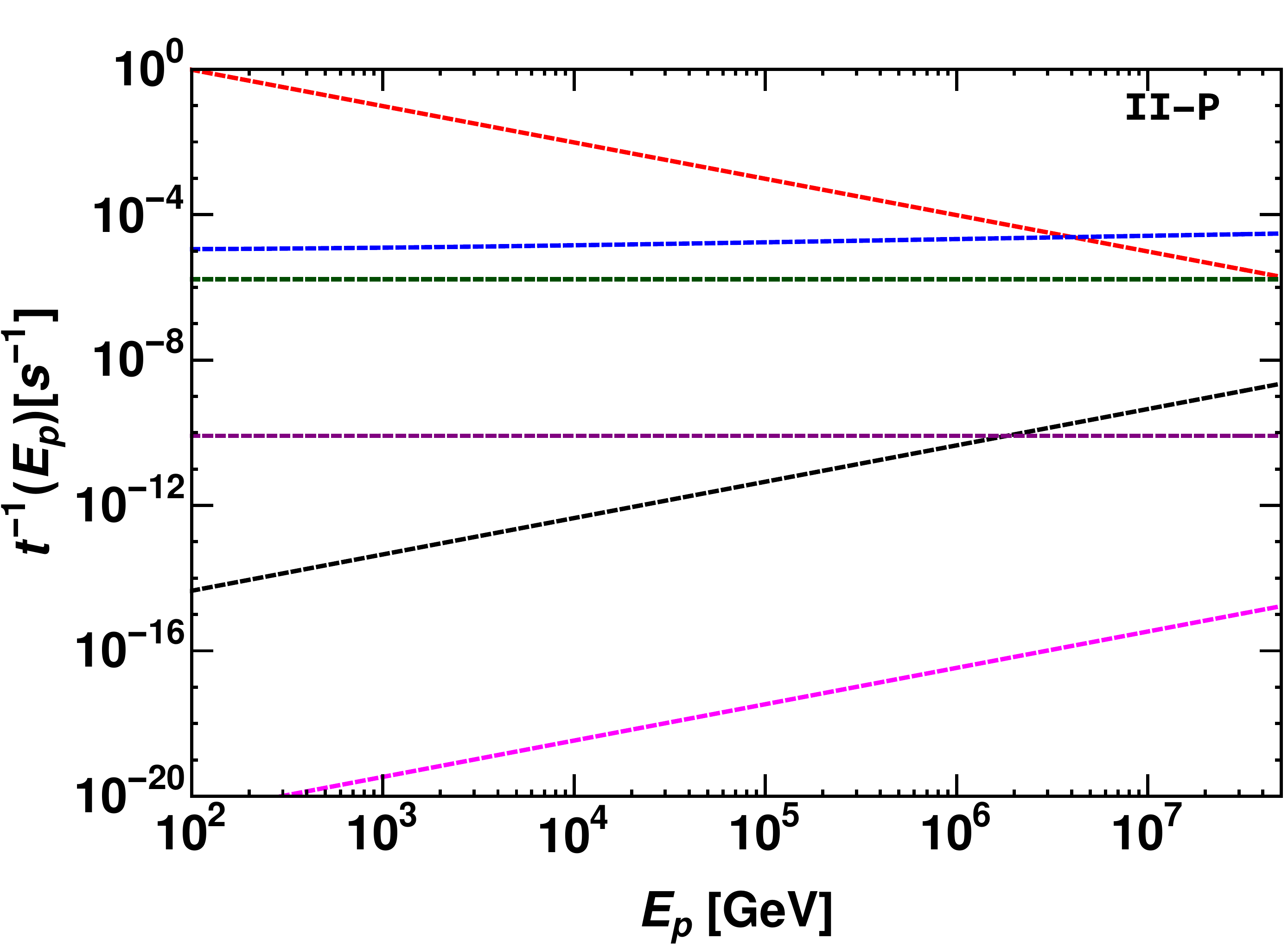}}
      }
    \caption{Acceleration time scale of protons and various proton energy loss processes as functions of the proton energy for  Type IIn YSNe (top panels) and Type II-P SNe (bottom panels) using the parameters in \ref{tab:parameters} and Sec. \ref{sec:DOYSNe}. 
    The left panels show the interaction time for $r=r_{\rm in}$, while right panels consider 
    $r=10\ r_{\rm in}$. Proton cooling is dominated by  $pp$ losses and therefore the total cooling timescale is the same as the $pp$ collision timescale.
    }
    \label{timescales}
\end{figure}

\subsection*{Acceleration time scale} The proton acceleration time scale is given by \cite{Protheroe:2003vc}:
\begin{equation}
   t_{\rm acc}= \frac{6 E_{\rm p}  c}{e B v_{\rm sh}^2}\ ,
\end{equation}
where the magnetic field $B$ is given in Sec.~\ref{sec:Model}.

\subsection*{Adiabatic loss time scales}
The adiabatic loss time scale for protons is same as the dynamical time scale and is given by: 
\begin{equation}
    t_{\rm ad} \sim t_{\rm dyn}= \frac{r}{v_{\rm sh}}.
\end{equation}

\subsection*{Proton-proton collision time scale}
The $pp$ collision time scale is given by
\begin{equation}
    t_{\rm pp}=(\kappa_{\rm pp}\sigma_{\rm pp} (E_{\rm p}) n_{\rm CSM}(r) c)^{-1},
\end{equation}
where $\kappa_{\rm pp}=0.5$ is the $pp$ collision inelasticity.

\subsection*{Proton synchrotron time scale}
High energy protons may lose energy due to proton synchrotron radiation. The corresponding synchrotron time scale is given by~\cite{padmanabhan_2000,Ghisellini:2019lgz}:
\begin{equation}
    t_{\rm syn}=\frac{\gamma m_{\rm p} c^2}{P_{\rm syn}},
\end{equation}
where $\gamma=E_{\rm p}/m_{\rm p} c^2$ and $P_{\rm{syn}}$ is the synchrotron power loss and is given by
\begin{equation}
    P_{\rm syn}= \frac{4}{3} \sigma_{T} c \left(\frac{m_{\rm e}}{m_{\rm p}}\right)^2 \frac{B^2(r)}{8 \pi} \left(\frac{v_{\rm sh}}{c}\right)^2\gamma^2\ ,
\end{equation}
where $\sigma_{\rm T}$ is the Thomson scattering cross-section.

\subsection*{Inverse Compton loss time scales}
Protons may lose energy to low energy SN photons via inverse Compton scattering and the corresponding loss time scale is given by~\cite{padmanabhan_2000}
\begin{equation}
    t_{\rm IC}=\frac{\gamma m_{\rm p} c^2}{P_{\rm IC}},
\end{equation}
where the inverse Compton power $P_{\rm IC}$ is given by
\begin{equation}
    P_{\rm IC} = \frac{4}{3} \sigma_{\rm T} c U_{\rm ph}(r) \gamma^2 \left(\frac{v_{\rm sh}}{c}\right)^2\left(\frac{m_{\rm e}}{m_{\rm p}}\right)^2\ ,
\end{equation}
where, $U_{\rm ph}(r)$ is the energy density of photons in CSM~\cite{Petropoulou:2016zar}:
\begin{equation}
    U_{\rm ph}(r)=\frac{{L_{\rm SN,pk}}}{{(4 \pi c r_{\rm in}^2)}} \left(\frac{r_{\rm in}}{r}\right)^2.
    \label{th_photon}
\end{equation}
The SN peak luminosity $L_{\rm SN,pk}$ for different YSNe can be found in Sec.~\ref{sec:DOYSNe}.

\subsection*{Bethe-Heitler energy loss time scale}
The Bethe-Heitler energy loss time scale is

\begin{equation}
    t_{\rm BH}=\tau_{\rm BH} \left(\frac{r}{r_{\rm in}}\right)^2
\end{equation}
where $\tau_{\rm BH}^{-1}={(\kappa_{\rm BH}\sigma_{\rm BH}c U_{\rm ph}(r_{\rm in}))}/{\epsilon_{\rm av}}$ and $\kappa_{\rm BH}\sigma_{\rm BH}\approx 6 \times 10^{-31}$ $\rm cm^2$~\cite{Petropoulou:2016zar}.

\subsection*{Dominant energy loss processes}
Fig.~\ref{timescales} shows the   time scales introduced above as functions of the proton energy for the dominant YSN classes: SNe IIn (top panel) and II-P (bottom panel), see  Table~\ref{tab:parameters}. Since the CSM density and SN thermal photon density change with the shock radius, the acceleration timescale and different cooling timescales also change. Therefore, we show the evolution of these timescales for two different values of shock radius, i.e., $r=r_{\rm in}$ (left panel) and $r=10r_{\rm in}$ (right panel). 
For both classes of SNe, the dominant energy loss processes are  $pp$ collisions and adiabatic losses for both shock radii. Note that  photo-hadronic  ($p\gamma$) interactions are not important here as the average energy of SN thermal photons is about $1$~eV; the threshold energy of protons for this process is   $\sim 10$~PeV and the SN flux of protons above this energy is very small. Moreover, the cross-section of $p\gamma$ interactions ($\sim 10^{-28}$ $\rm cm^{2}$~\cite{Kelner:2008ke}) is much smaller than that of $pp$ interaction ($\sim 10^{-26}$ $\rm cm^{2}$~\cite{Kelner:2006tc}).


\bibliographystyle{JHEP}
\bibliography{draft-JCAP.bib}

\end{document}